\documentclass[aps,
twocolumn,nofootinbib]{revtex4-1}

\usepackage{url,comment}
\usepackage{times}
\usepackage{latexsym}
\usepackage{graphicx, graphics, hyperref, amsmath, amssymb, slashed, xcolor, bbm,amsthm, array}
 \usepackage{subfigure}
\usepackage{rotating}
\usepackage{afterpage}
\newcommand{\nc}{\newcommand}
\nc{\beq}{\begin{equation}}
\nc{\eeq}{\end{equation}}
\nc{\barray}{\begin{eqnarray}}
\nc{\earray}{\end{eqnarray}}
\nc{\barrayn}{\begin{eqnarray*}}
\nc{\earrayn}{\end{eqnarray*}}
\nc{\bcenter}{\begin{center}}
\nc{\ecenter}{\end{center}}
\nc{\mc}{\mathcal}
\nc{\er}[1]{(\ref{eq:#1})}
\nc{\onehalf}{\frac{1}{2}} 
\nc{\partialbar}{\bar{\partial}}
\nc{\psit}{\widetilde{\psi}}
\nc{\Tr}{\mbox{Tr}}
\nc{\hc}{\mbox{H.c.}}
\nc{\ev}{\;\mathrm{eV}}
\nc{\mev}{\;\mathrm{MeV}}
\nc{\gev}{\;\mathrm{GeV}}
\nc{\tev}{\;\mathrm{TeV}}

\def\chii0{\chi_i^0}
\def\chij0{\chi_j^0}

\newcommand{\gsim}{\lower.7ex\hbox{$\;\stackrel{\textstyle>}{\sim}\;$}}
\newcommand{\lsim}{\lower.7ex\hbox{$\;\stackrel{\textstyle<}{\sim}\;$}}
\nc{\ttbar}{t\bar t}
\def\ifb{{\ \rm fb}^{-1}}
\def\iab{{\ \rm ab}^{-1}}
\def\ipb{{\ \rm pb}^{-1}}

\newcommand{\fref}[1]{Fig.~\ref{f.#1}}

\newcommand{\sref}[1]{Section~\ref{s.#1}}

\newcommand{\cref}[1]{Chapter~\ref{c.#1}}

\graphicspath{{figs/}}


\begin{document}
\preprint{TTP17-053}

\title{
New Physics Opportunities for Long-Lived Particles at Electron-Proton Colliders
}

\author{David Curtin}
\email{dcurtin1@umd.edu}

\author{Kaustubh Deshpande}
\email{ksd@umd.edu}
\affiliation{Maryland Center for Fundamental Physics, Department of Physics,\\ University of Maryland, College Park, MD 20742-4111 USA}

\author{Oliver Fischer}
\email{oliver.fischer@kit.edu}
\affiliation{Institute for Nuclear Physics (IKP), Karlsruhe Institute of Technology, 
Hermann-von-Helmholtz-Platz 1, D-76344 Eggenstein-Leopoldshafen, Germany}

\author{Jos\'e Zurita }
\email{jose.zurita@kit.edu}

\affiliation{
\mbox{Institute for Nuclear Physics (IKP), Karlsruhe Institute of Technology,}\\
\mbox{Hermann-von-Helmholtz-Platz 1, D-76344 Eggenstein-Leopoldshafen, Germany}\\ \phantom{a}\vspace*{-2mm} \\
\mbox{Institute for Theoretical Particle Physics (TTP), Karlsruhe Institute of Technology,}\\ 
\mbox{Engesserstraße 7, D-76128 Karlsruhe, Germany} \vspace*{2mm}
}

\date{\today}
\begin{abstract}
Future electron-proton collider proposals like the LHeC or the FCC-eh can supply $\iab$ of collisions with a center-of-mass energy in the TeV range, while maintaining a clean experimental environment more commonly associated with lepton colliders.
We point out that this makes $e^- p$ colliders ideally suited to probe BSM signatures with final states that look like ``hadronic noise'' in the high-energy, pile-up-rich environment of $pp$ colliders. 
We focus on the generic vector boson fusion production mechanism, which is available for all BSM particles with electroweak charges at mass scales far above the reach of most lepton colliders.
This is in contrast to previous BSM studies at these machines, which focused on BSM processes with large production rates from the asymmetric initial state. 
We propose to exploit the unique experimental environment in the search for long-lived particle signals arising from Higgsinos or exotic Higgs decays. 
At $e^- p$ colliders, the soft decay products of long-lived Higgsinos can be explicitly reconstructed  (``displaced single pion''), and very short lifetimes can be probed. 
We find that $e^- p$ colliders can explore significant regions of BSM parameter space inaccessible to other collider searches, with important implications for the design of such machines. 

\end{abstract}



\maketitle


\section{Introduction}
\label{s.intro}

Progress in high energy physics relies on designing new experiments to explore ever higher mass scales and smaller interactions~\cite{Quigg:2017moa}. 
This is vital both to understand the Standard Model (SM) at new energy regimes, as well as for the discovery of Beyond SM (BSM) physics.
As the Large Hadron Collider (LHC) makes impressive progress exploring of the TeV scale, it is therefore a high priority to look ahead and identify the most important physics opportunities presented by the next round of proton and electron colliders. 

Lessons learned from the LHC era provide important context for any future collider program (see e.g.\ ref.~\cite{Ellis:2017avv}). 
When the LHC experiment was designed more than two decades ago, the main focus was the discovery of the Higgs boson and searches for BSM theories like supersymmetry (SUSY) ~\cite{Martin:1997ns}. 
This meant that identification of high energy final states, copiously produced in prompt decays of intermediate particles with masses around the TeV scale, was paramount.
The exploration of this canonical ``High Energy Frontier'' will be an important goal for future experiments, but the absence (to date) of any such BSM signatures at the LHC presents us with an important puzzle: 
How do we reconcile LHC null results with the fact that motivation for BSM theories is as strong as ever? 
The hierarchy problem has been sharpened by the discovery of the Higgs and explicitly calls for TeV-scale new physics, while dark matter, baryogenesis and neutrino masses continue to beg for explanations. 
An important lesson of the last decade is that these fundamental mysteries can be addressed by theories which have signatures very unlike the high energy SUSY signals of the canonical high energy frontier. Hidden valleys~\cite{Strassler:2006im,Strassler:2006ri,Strassler:2006qa,Han:2007ae,Strassler:2008bv,Strassler:2008fv}, Hidden Sectors connected to Dark Matter~\cite{Baumgart:2009tn, Kaplan:2009ag,
  Chan:2011aa, Dienes:2011ja, Dienes:2012yz, Kim:2013ivd}, Neutral Naturalness~\cite{Burdman:2006tz, Cai:2008au, Chacko:2005pe}, WIMP baryogenesis~\cite{Cui:2012jh, Barry:2013nva, Cui:2014twa, Ipek:2016bpf}, many varieties of SUSY \cite{Arvanitaki:2012ps,ArkaniHamed:2012gw,Giudice:1998bp,Barbier:2004ez,Csaki:2013jza,Fan:2011yu}, and right-handed neutrinos~\cite{Helo:2013esa, Antusch:2016vyf,Graesser:2007yj, Graesser:2007pc, Maiezza:2015lza, Batell:2016zod,Blondel:2014bra} might only show up in ``exotic channels'' like Long-Lived Particle (LLP) signatures. 
It is important that future colliders can explore this ``Lifetime Frontier" as well as the High Energy or High Intensity Frontiers. 

{\bf Future colliders:} Most proposals fall into two categories: lepton or hadron colliders. 
The proposed $e^+ e^-$ colliders, namely the ILC in Japan  \cite{Baer:2013cma,Brau:2015ppa}, the CEPC in China \cite{CEPC-SPPCStudyGroup:2015csa}, and the FCC-ee (formerly known as TLEP) \cite{Gomez-Ceballos:2013zzn} and CLIC at CERN \cite{Aicheler:2012bya} are ideal for precision measurements of the Higgs boson properties due to their exquisitely clean experimental environment.
The sensitivity of the Higgs to the existence of new physics (see e.g.~\cite{Curtin:2013fra}) makes this an endeavor of the highest priority, but direct discovery of new BSM states at such machines is generally less likely, since their center of mass energy is below that of the present LHC.

On the other hand, presently discussed future $p p$ colliders like the FCC-hh at CERN~\cite{Golling:2016gvc,Mangano:2016jyj,Contino:2016spe} or the SppC in China~\cite{Tang:2015qga} would offer enormous center of mass energies at the 100 TeV scale as well as huge event rates for many weak-scale processes like Higgs Boson production. This would enable them to probe very high mass scales and very rare processes, provided the final states can be identified in such an extremely high-energy high-rate environment. 

There is a hybrid of these two approaches which is less often discussed: electron-proton colliders. HERA was the only such machine ever built, and it was instrumental to establish the inner structure of the proton via deep inelastic scattering (DIS) measurements.
The resulting information about Parton Distribution Functions (PDFs) is now part of textbooks and Monte Carlo generators.
This was HERA's primary objective, and its successes are of foundational importance for high energy measurements and BSM searches at $pp$ colliders like the Tevatron and the LHC. 
HERA's direct contributions to BSM searches, however, were much more limited. 
The electron-proton initial state does not give rise to large cross sections for many BSM processes, and HERA's center-of-mass energy of $\sqrt{s} = 320 \gev$ and integrated luminosity of $\sim 500 \ipb$  was far below the Tevatron's $1.96 \tev$ and $10 \ifb$. As a result, HERA was outclassed in mass reach for almost all BSM signatures, with the exception of some leptoquark scenarios~\cite{South:2013fta, Abazov:2011qj}. 

{\bf Beyond HERA:}
Plans for electron-proton colliders have evolved considerably since HERA. 
Modern proposals envision them  an ``add-on'' or ``upgrade'' to an existing high-energy $pp$ collider, at a cost that is roughly an order of magnitude below that of the $pp$ machine alone.
The LHeC proposal \cite{Klein:2009qt,AbelleiraFernandez:2012cc,Bruening:2013bga} consists of a 60 GeV high-intensity linac supplying the electron beam to meet the 7 TeV proton beam at a collision point in the LHC tunnel. This includes a dedicated detector, with a geometry that accommodates the asymmetric nature of the collision.
The LHeC would have a center of mass energy of $1.3 \tev$ and is planned to deliver up to $1 \iab$ of collisions over its approximately 10-year lifetime, a drastic increase of energy and especially luminosity compared to HERA.
An analogous proposal, FCC-eh, exists for a future 100 TeV $pp$ collider at CERN \cite{Zimmermann:2014qxa}, but one could just as easily imagine such an extension for the HE-LHC~\cite{Bruning:2011dm} or the SppC~\cite{Tang:2015qga}.

Future machines like the LHeC or the FCC-eh would  greatly advance our knowledge of the proton~\cite{Klein:2016uwv} with many important benefits for the main $pp$ program,  but the physics potential does not stop there. 
Future $e^- p$ machines can access mass scales beyond the energies of  lepton colliders, while maintaining a clean experimental environment and delivering high luminosity, all for a fraction of the cost.
This explains their perhaps surprising ability to support a strong precision Higgs program~\cite{utaklein_fccweekberlin,Tang:2015uha,Kumar:2015kca,Kumar:2015tua,Zhang:2015ado}: LHeC measurements of Higgs couplings relying on Vector Boson Fusion (VBF) production might be competitive with electron colliders (albeit without the important model-independent measurement of the Higgs width via $Zh$ production).

Could we harness this unique experimental setup to explore hitherto inaccessible BSM signatures as well?
Previous studies exploring the BSM reach of future $e^-p$ colliders mostly focused on \emph{production modes} that allowed for large signal rates from the asymmetric initial state: leptoquarks~\cite{AbelleiraFernandez:2012cc},  4$^\mathrm{th}$ generation quarks \cite{Cakir:2009xi}  or excited leptons \cite{Liang:2010gm}, 
right-handed (RH) neutrinos \cite{Liang:2010gm, Blaksley:2011ey, Mondal:2016kof, Duarte:2014zea, Antusch:2016ejd}, and left-right symmetric models with new gauge bosons in the $t$-channel \cite{Mondal:2015zba, Lindner:2016lxq}. However, in all of those cases, with the exception of RH neutrino models (which include LLP signals \cite{Antusch:2016ejd}), the LHC or HL-LHC has higher mass reach~\cite{Diaz:2017lit, Dorsner:2016wpm, Raj:2016aky, Khachatryan:2015scf, Aad:2016shx, Aad:2015voa, Ferrari:2000sp}. This is a familiar echo of the HERA-Tevatron interplay. 
One might think na\"ively that this puts a damper on the BSM motivation for electron-proton colliders, but we argue that this conclusion is premature. 

In fact, we argue that $e^- p$ colliders are uniquely suited to discover new physics, with strengths that are truly complementary to both  $pp$ and $e^+ e^-$ programs. Given the unknown nature of new physics signatures in light of the LHC puzzle, this makes $e^- p$ colliders a vital component of a future high energy physics program. 

{\bf Focusing on the final state:}
Rather than focusing on BSM scenarios with large production rates, we suggest focusing on BSM scenarios which give rise to \emph{final states that look like hadronic noise} in the pile-up-rich environment of $pp$ colliders. 
The clean environment of the $e^- p$ collider allows for their unambiguous reconstruction, while their large center-of-mass energies allow them to access higher mass scales than lepton colliders. 
This view is tentatively backed up by the encouraging results of the initial precision Higgs and RH neutrino studies, which relied heavily on the clean experimental environment.
The shifted focus from initial to the final state also allows us to consider more general BSM production modes like VBF, which are present in any theory with new electroweak charged states. 
We consider LLP signatures to demonstrate the utility of this new paradigm.

{\bf Long lived particles:}
New states with macroscopic lifetime are extremely broadly motivated. 
They often emerge as result from basic symmetry principles of Quantum Field Theory and are highly generic in BSM theories, where states can be long-lived due to approximate symmetries, modest mass hierarchies, or sequestration of different sectors in a UV completion. 
As outlined above, they are ubiquitous in theories of hidden valleys and general hidden sectors, and are the smoking gun signal of Neutral Naturalness, certain varieties of SUSY, theories explaining the origin of neutrino masses, as well as many baryogenesis and dark matter scenarios.

LLPs can be detected directly via their passage through the  detector material if they are charged or colored (and long-lived enough), or by reconstruction of a \emph{displaced vertex} (DV) if they decay in the detector. 
They are not picked up by most standard searches focusing on prompt signals, making them consistent with recent LHC null results. 
However, the spectacular nature of these signals means that \emph{dedicated} LLP searches typically have very low backgrounds, often allowing for discovery with just a few observed events at the LHC or future colliders~\cite{CMS:2014hka,ATLAS:2016jza,CMS:2014wda,Aad:2015uaa,Blondel:2014bra,Antusch:2016vyf,Antusch:2016ejd}
There are, however, important regions of LLP signature space which are very difficult for $pp$ colliders to probe, due to low signal acceptance, trigger thresholds, or sizable backgrounds. 
This includes (i)~invisible LLPs with very long lifetimes that escape the main detectors, (ii) LLPs with very soft decay products, and (iii) LLPs with very short lifetimes $\lesssim$ mm, making them difficult to distinguish from hadronic backgrounds. 
Recent proposals for dedicated external LLP detectors near an LHC collision point, like MATHUSLA~\cite{Chou:2016lxi,Curtin:2017izq}, milliQan~\cite{Ball:2016zrp}, CODEX-b~\cite{Gligorov:2017nwh} and FASER~\cite{Feng:2017uoz}, aim to address the first of these shortcomings.
The second and third class of signals are prime targets for $e^- p$ colliders.

We examine two important BSM signatures at $e^-p$ colliders after briefly reviewing the salient details of these proposals in \sref{epreview}. 
We study Higgsinos in \sref{higgsinos}. If the winos are decoupled, the charged Higgsino can have a lifetime of up to several mm, decaying to often just a single soft pion via a small mass splitting to the neutral Higgsino. This decay cannot be reconstructed at $pp$ colliders, forcing searches to rely on monojet or disappearing track signals. In the clean environment of $e^- p$ colliders, these soft displaced final states can be explicitly reconstructed, and lifetimes many orders of magnitude shorter than those accessible by $pp$ colliders can be probed at masses far beyond the reach of lepton colliders. 
To demonstrate the utility of $e^-p$ colliders for general LLP signals with very short lifetime, we also consider LLP production in exotic Higgs decays in \sref{higgs}. Again, the $e^- p$ searches outperform searches for $pp$ colliders by orders of magnitude for very short lifetimes. 
We conclude in \sref{conclusion}.


\section{Electron-proton collider basics}
\label{s.epreview}

\begin{figure}[t]
\begin{center}
\includegraphics[width=0.4 \textwidth]{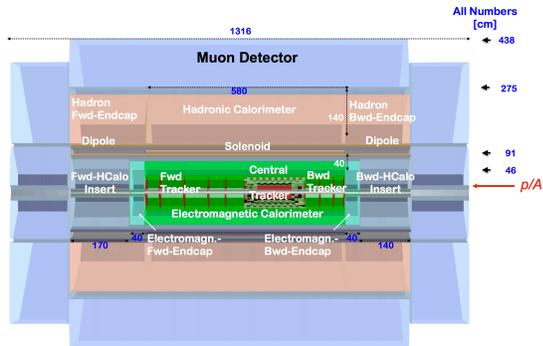}
\end{center}
\caption{
Possible layout of the LHeC detector, figure from~\cite{AbelleiraFernandez:2012cc}.
}
\label{f.ep}
\vspace*{-3mm}
\end{figure}

Electron-proton colliders are hybrids between $e^-e^+$ and $pp$ colliders. 
Today's proposals consider electron beams from a linac that intersect with the hadron beam from an existing $pp$ collider (though using an electron beam from a circular collider would also be possible). 
Such machines allow for a clean collision environment with very little pile-up, center-of-mass energies of ${\cal O}(1) \tev$  and luminosities of $1 \iab$ or more.
 
The Large Hadron electron Collider (LHeC) \cite{Klein:2009qt,AbelleiraFernandez:2012cc,Bruening:2013bga} is a proposed upgrade for the high luminosity phase of the LHC. 
It foresees the construction of a high-intensity electron accelerator adjacent to the main rings. 
The resulting 60 GeV $e^-$ beam would meet the 7 TeV proton beam from the LHC at a dedicated interaction point in the HL-LHC tunnel, with an envisaged total luminosity of $1 \iab$ at a 1.3 TeV center-of-mass energy over the lifetime of the program. We remark that higher electron beam energies are also discussed~\cite{AbelleiraFernandez:2012cc}.
The collisions would be analyzed in a general-purpose detector, with an adjusted geometry to accommodate the asymmetric collision.

An even more powerful electron-proton collider is discussed as part of the Future Circular Collider design study, namely the Future Circular electron-hadron Collider (FCC-eh) \cite{Zimmermann:2014qxa}. 
The FCC-eh is based on the electron beam from the LHeC facility, colliding with the 50 TeV proton beam from the hadron-hadron mode of the FCC. The final integrated luminosity is currently assumed to be $\sim 1$ ab$^{-1}$, at center-of-mass energies up to 3.5 TeV \cite{Klein:2016uwv}. In the following, we will refer to this experimental setup as the FCC-eh (60) to indicate the electron beam energy.

The goal of our study is to assess the BSM potential of $e^- p$ colliders, which should be a major design driver for the electron accelerator and detector. 
The FCC-eh specifications are much less finalized than the LHeC, and  it is  instructive to consider alternatives to the existing proposal, and how they differ in BSM reach. 
We will therefore also discuss a version of the FCC-eh which represents perhaps the highest-energy setup that might be realistic: an electron beam with energy 240 GeV meeting the 50 TeV proton beam, to generate center-of-mass energies of $6.9$ TeV. We refer to this scenario as the FCC-eh (240).
Such a high energy electron beam would be challenging to implement, but there are several options, including a nearby ILC or CLIC-like facility or use of a high-energy circular electron-positron collider in the same tunnel (as is planned in the CEPC/SppC project in China).\footnote{In the context of the FCC-ee, the maximum energy that may be feasible from a {technological point of view} is $\sim$ 250 GeV \cite{Zimmermann:2014qxa}.} 
Morevoer, since the benchmark luminosity of the FCC-hh program is $\sim$10 times higher than foreseen for the HL-LHC, we also consider the analogous possibility of $10 \iab$ at the FCC-eh (60) and FCC-eh (240).

The LHeC detector layout from the technical design report is shown in \fref{ep}~\cite{AbelleiraFernandez:2012cc}. 
Precise details of the detector are not relevant for our benchmark studies, and we only focus on the most salient features. For concreteness, and also to be somewhat conservative, we assume the same detector capabilities for the FCC-eh as for the LHeC (though this does not affect our qualitative conclusions). 

Notable is the tracker coverage to very high rapidity of 4.7 in the forward and backward direction with respect to the proton beam, starting at a distance of about 3cm from the beams.
The detector has a magnetic field of $\sim 3.5$ T, and the nominal tracking resolution is 8 $\mu$m. Studies for ILC detectors show that impact parameter resolutions down to  $\sim$5 $\mu$m may be possible \cite{Brau:2007zzb,Behnke:2013lya,vanderKolk:2017urn}.
To assess the importance of tracking resolution on LLP reach, we therefore consider resolutions of 5, 8 and 16 $\mu$m.
The elliptical interaction point has rms dimensions of 7 $\mu$m in the transverse plane and 0.6 mm along the longitudinal beam direction. 
Charged hadronic tracks with energies above few GeV are generally accepted by the calorimeters.
However, since we will be considering LLPs that decay to soft low-multiplicity hadrons, precise energy thresholds will be important.
To assess their impact on LLP reach we consider  $p_T$ thresholds of 50, 100 and 400 MeV for reliable reconstruction of a single charged particle track. 
The trigger capabilities of the tracking system are not yet completely defined~\cite{AbelleiraFernandez:2012cc}, but since DIS measurements are a major design driver, we assume that single jets with $p_T > 20 \gev$ can be triggered on with high efficiency. This means trigger considerations will not play a major role in our analyses.

With the above specified performance parameters, the corresponding $e^- p$ collider concepts offer center-of-mass energies larger than all but the most ambitious lepton collider proposals, while maintaining a very clean experimental environment.
In comparison to $pp$ colliders, the various hadronic backgrounds have very different distributions and are strongly suppressed. 
At the LHeC, the pile-up is expected to be $\sim 0.1$ per event, while for the FCC-eh (60) it may rise to $\sim 1$. 
We will consider analysis strategies which take advantage of, but are robust with respect to, these low pile-up levels.

\section{Long-lived Higgsinos}
\label{s.higgsinos}

The electroweakinos (EWinos) of the MSSM are well-motivated candidates for LLPs.
The mixing of the Bino, Wino and Higgsino fields gives rise to four neutralino and two chargino mass eigenstates. 

If the mixing of these particles is significant they can be detected at hadron colliders via searches for high energy leptons and missing energy \cite{Aad:2014vma,Sirunyan:2017lae}.

In the following we consider the challenging limit of small mixing. 
In that case, the masses of the lightest Higgsino (Wino) chargino and the lightest neutralino are only slightly split due to electroweak symmetry breaking loop effects.\footnote{These cases are often referred to in the literature as `pure' limits. We note that a `pure Bino' that is stable on cosmological time scales and thus a viable dark matter candidate needs to be lighter than 100 GeV not to overclose the universe, which is ruled out by LEP searches\cite{Patrignani:2016xqp}.}
The difference between these two masses, referred to as the `mass splitting' ($\Delta m$) in the following, is  $\mathcal{O}(100) \mev$ which corresponds to a lifetime $c \tau \sim 7 \mathrm{mm}$ ($\sim$ 6 cm). 
Charged LLPs with this lifetime, decaying into a massive neutral particle, can be searched for at the LHC via so-called `disappearing-track searches'. 
Owing to the larger lifetime and four times larger production cross section,\footnote{The Casimir group factor is given simply by $T_3^2$.} Wino searches have significant mass reach at the LHC and FCC-hh \cite{Low:2014cba,Cirelli:2014dsa}.
Searches for Higgsinos are much more challenging, and a customized tracker with sensitivity to shorter lifetimes is needed, as shown in ref.~\cite{Mahbubani:2017gjh} (see also ref.~\cite{Fukuda:2017jmk}). 
Due to the almost-degenerate mass spectrum, the leptons and jets from the chargino decay have very small momenta and thus largely fail to pass reconstruction thresholds of the LHC analyses. Depending on the value of $\Delta m$, searches that include an ISR jet and additional `soft' leptons can yield relevant constraints \cite{Han:2013kza,Gori:2013ala,Schwaller:2013baa,Han:2014kaa,Baer:2014cua,Baer:2014kya,Badziak:2015qca,Han:2015lma,Barducci:2015ffa}. 
In scenarios where the mass splitting of the electroweakinos is given by the loop effects only, the relevant signature at the LHC is the missing energy, which is included in the so-called mono-jet searches.

There are important incentives to study Higgsino signatures beyond their role in supersymmetry. 
Neutral Higgsinos are thermal DM relics that can yield the observed relic density if their masses $m_\chi$ is around $1.1 \tev$ \cite{ArkaniHamed:2006mb} or below (depending on mixing). 
Furthermore, the lessons learned from studying pure Higgsinos can easily be transferred to theories with similar phenomenology, for instance models with inert multiplets \cite{LopezHonorez:2006gr,Cirelli:2005uq,Fischer:2013hwa} and vector-like leptons (see e.g.~\cite{Fujikawa:1994we,Joglekar:2012vc,ArkaniHamed:2012kq,Joglekar:2013zya,Falkowski:2013jya,Kumar:2015tna}), which are also interesting in the context of minimal models for gauge unification~\cite{EmmanuelCosta:2005nh, Bhattacherjee:2017cxh}. 
This makes the `pure-Higgsino' case very theoretically compelling, even as their low production cross section, soft decay products, and short lifetime make them the most experimentally challenging electroweakino scenario at proton-proton colliders. 

In the remainder of this section we review the main phenomenological features, branching ratios and lifetimes of Higgsinos. After setting the stage by summarizing current and projected constraints from cosmology and $pp$ colliders, we  show how $e^- p$ colliders can fill in crucial gaps in coverage.

\subsection{Higgsino Phenomenology}

The spectrum and interactions of EWinos in the MSSM has been studied in depth \cite{Haber:1984rc,Martin:1997ns}, and we only focus on the aspects relevant for our analysis here. 
In the decoupled Wino limit where $\mu \ll M_2$ and $\mu < M_1$ there is one charged state $\chi^{\pm}$ and three neutral $\chi_i^0, i =1,2,3$. The mass of the charged state receives the 1-loop correction from EW gauge bosons,  $\Delta_\mathrm{1-loop}$. In the neutral sector the two lighter states are at about the scale $\mu$ split by $\Delta_0$ and the third one at the heavy scale $M_1$. The latter does not impact directly on the phenomenology, but rather dictates $\Delta_0$. One can thus trade the Lagragian parameters $\mu, M_1, \tan \beta$ for the mass of the lightest neutralino $m_{\chi^0_1}$ and the mass splitting with respect to the chargino ($\Delta m \equiv m_{\chi^\pm} - m_{\chi^0_1})$ and to the second neutralino ($\Delta_0 \equiv m_{\chi^0_2} - m_{\chi^0_1}$). The relevant expressions read
\begin{eqnarray}
\nonumber  m_{\chi^0_1} &=& | \mu | - \frac{m^2 (1 + \mathrm{sign}(\mu) s_{2\beta})}{2 M_1 (1 - |\mu| / M_1) } \, ,\\
\Delta m 
&=& \Delta_\mathrm{1-loop} + \frac{m^2 (1 + \mathrm{sign}(\mu) s_{2\beta})}{2 (M_1 - |\mu|)} \, ,\\
\Delta_0 
\nonumber &=& \frac{m^2}{M_1} \bigl(\frac{1+ \mathrm{sign}(\mu)  s_{2\beta} \mu/M_1 }{1-\mu^2/M_1^2} \bigr) \, ,
\end{eqnarray}
where  $\tan \beta = v_u/v_d$, and the above results assume $m = m_Z s_W \approx 44 \gev \ll |M_1 - \mu|$. We consider $M_1$ to be real and positive, while $\mu$ is real with either sign. $\Delta_\mathrm{1-loop} \sim 300 \mev$  has very modest dependence on $m_{\chi^\pm}$, and one can see from the above expressions that  the dependence on $\tan \beta$ is modest as well. For concreteness, we take in our analysis  $\tan \beta = 15$.
The choice of $m_{\chi^\pm}$ and $\Delta m$ then determines the spectrum. Note that $\Delta m = m_{\chi^\pm} - m_{\chi^0_1} > \Delta_\mathrm{1-loop} > m_{\chi^\pm} - m_{\chi^0_2} $. 
Upscattering in direct detection experiments 
\cite{TuckerSmith:2001hy,TuckerSmith:2004jv} forces $\Delta_0 \gtrsim 0.1 \mev $, which implies an upper bound on $M_1 \lesssim 20 $ PeV.  

The neutralino couplings to the gauge bosons follow from the EW charges. 
The three particles with masses $\sim |\mu|$ are `almost-doublets', and hence the $Z$-current couples  $\chi_1^0$ and $\chi_2^0$ with 'almost-full' strength. Both the Z and Higgs interactions with the DM candidate $\chi_1^0$ arise from doublet-singlet mixing, and hence they are suppressed by powers of $m_Z /  |\mu|, m_Z / M_1$, which also suppresses the direct detection cross section, see section~\ref{ss.comple} below.

The decay modes of the long-lived chargino are computed using the expressions in refs.~\cite{Chen:1996ap,Chen:1999yf} and  shown in \fref{higgsinodecays}. Chargino decays to $\chi^0_1$ are always allowed with a mass splitting greater than $\Delta_\mathrm{1-loop}$, which sets the maximum possible lifetime in this model (though longer lifetimes can be considered in more general scenarios). If $M_1$ is much larger than $|\mu|$, the lifetime gets reduced by a factor of 2, as the chargino decays with a similar width to each neutralino. Note that this is unlike the Wino case, where there is only one neutralino in the low energy spectrum. For lower values of $M_1$, the chargino decays to $\chi^0_2$ become smaller.
The hadronic decay widths require some care due to the small mass splitting. 
For $\Delta m \lesssim 1 \gev$, one must compute partial widths to exclusive hadron final state like $\pi^+ \chi^0_1$. For $\Delta m \gg 1 \gev$, quarks are the relevant degrees of freedom, and hadronic decays give rise to jets which shower and hadronize.

In practice, we compute hadronic final states both in the exclusive hadron picture and the inclusive quark picture, and define $\Delta m_*$ as the mass splitting where $\sum \Gamma(\chi^\pm \to \mathrm{hadrons} + \chi^0_1) = \sum \Gamma(\chi^\pm \to \mathrm{quarks} + \chi^0_1)$. For $\Delta m < \Delta m_*$ we then use the hadron picture and for $\Delta m > \Delta m_*$ we use the quark picture, which is responsible for the sharp turn-over at $\Delta m \approx 1.75 \gev$ in \fref{higgsinodecays}. 
This unphysical sharp turn-over between the two regimes is sufficient at the level of detail of our study. 
To capture the effect of hadronization uncertainties, we follow ref.~\cite{Chen:1996ap} and compute the partial decay widths to quarks assuming $m_d = 0.5 \gev$ and $0 \gev$, with different $\Delta m_*$ for each case. 

We note a few important features of the branching ratios in \fref{higgsinodecays}. At small mass splitting, decays to both $\chi^0_1$ and $\chi^0_2$ are kinematically allowed while for larger mass splittings all decays are to  $\chi^0_1$. Our region of interest for displaced searches is $c \tau \gtrsim \mu m$, corresponding to $\Delta m \lesssim 2.5 \gev$. The branching fractions have some quantitative (but not qualitative) dependence on $\mathrm{sign}(\mu)$, but very little dependence on $m_{\chi^\pm}$ itself. As mentioned above, the minimal mass splitting is given by $\Delta_\mathrm{1-loop}$ and larger mass splittings are possible when $M_1$ is closer to $\mu$, although for our region of interest $M_1$ is still several TeV to tens of TeV.

On our scenario, LEP excludes $\chi^+$ masses below 104 GeV \cite{Patrignani:2016xqp}. The existing LHC searches for soft leptons \cite{Khachatryan:2015pot} are currently only sensitive to $\Delta \sim 20$ GeV. The prospects of the HL-LHC and of future colliders are summarized below.

\begin{figure}[t]
\begin{center}
\includegraphics[width=9cm]{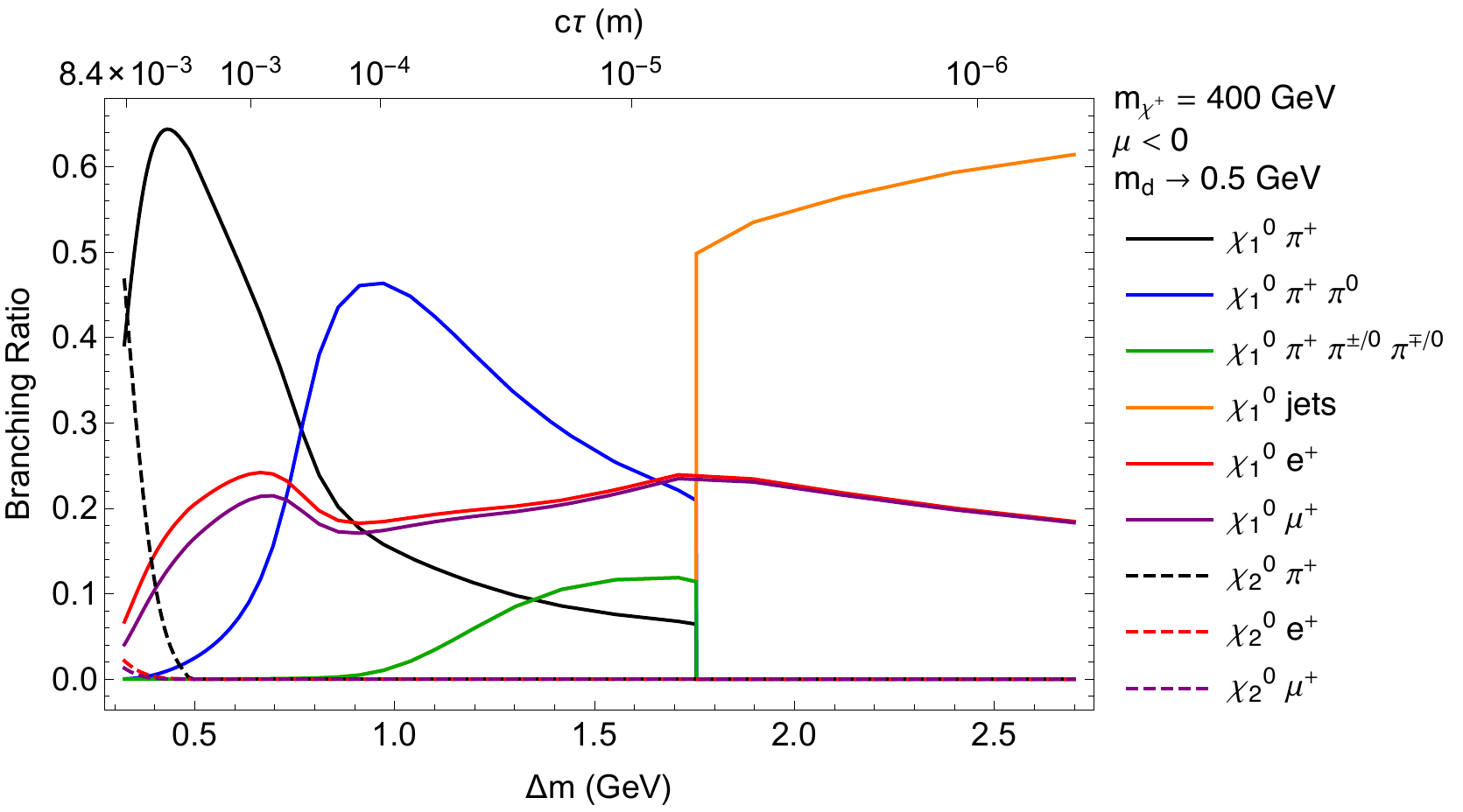}
\end{center}
\caption{
Decay branching ratios for a 400 GeV charged Higgsino as a function of $\Delta m = m_{\chi^\pm_1 - \chi^0_1}$ and $\mu < 0$. Note the chargino lifetime on the upper vertical axis.
Hadronic decay widths are computed assuming $m_d = 0.5 \gev$.
The switch from an exclusive hadronic final state description to an inclusive jet final state description 
occurs at around $\Delta m \approx 1.75 \gev$, which decreases to 1.3 GeV if the assumed $m_D$ is taken to zero. 
The $\mu > 0$ case is qualitatively very similar, and there is very little dependence on the Higgsino mass.
}
\label{f.higgsinodecays}
\end{figure}

\subsection{Probing Higgsinos with $pp$ colliders and cosmology}
\label{ss.comple}

\begin{figure}
\begin{center}
\includegraphics[width=0.5 \textwidth]{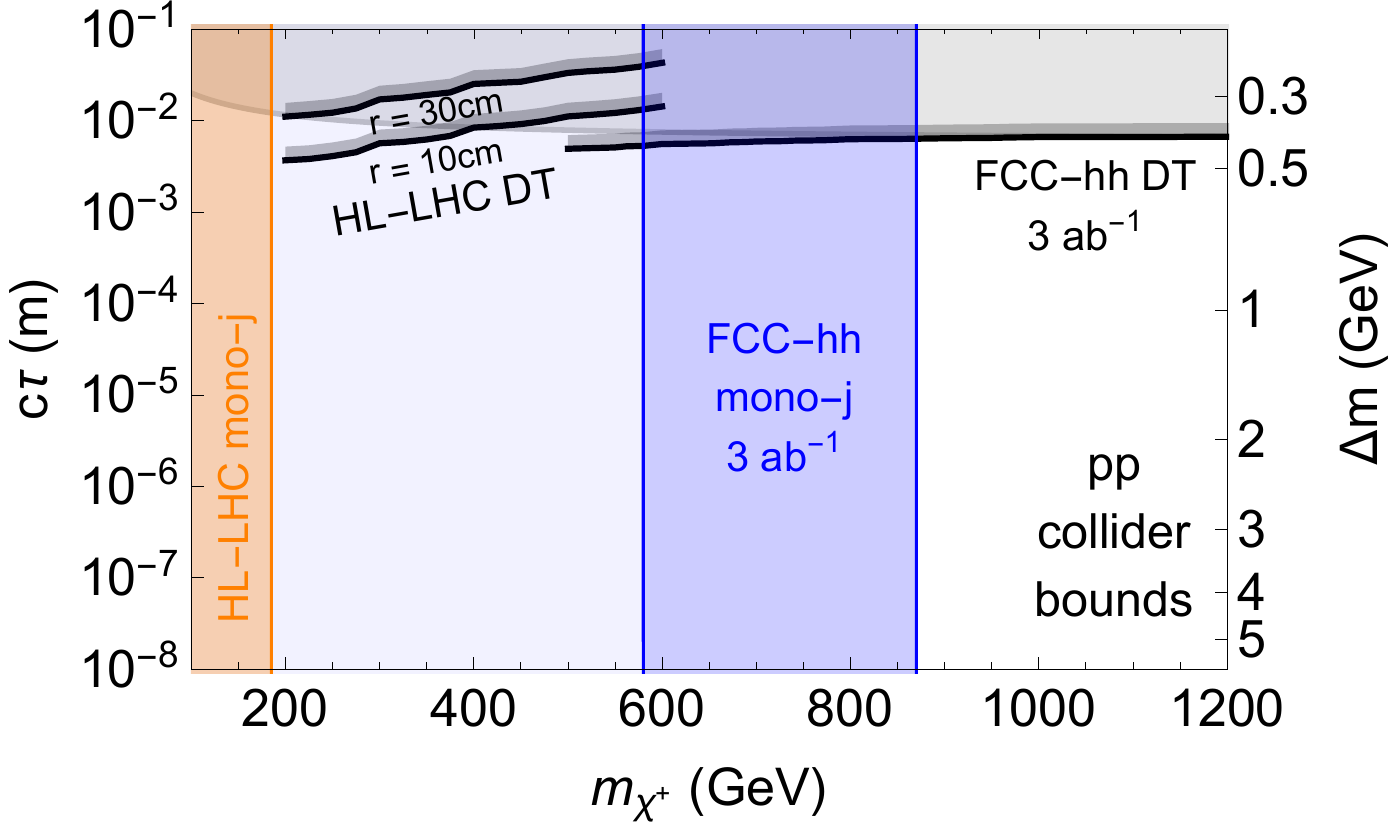}
\\
\includegraphics[width=0.5 \textwidth]{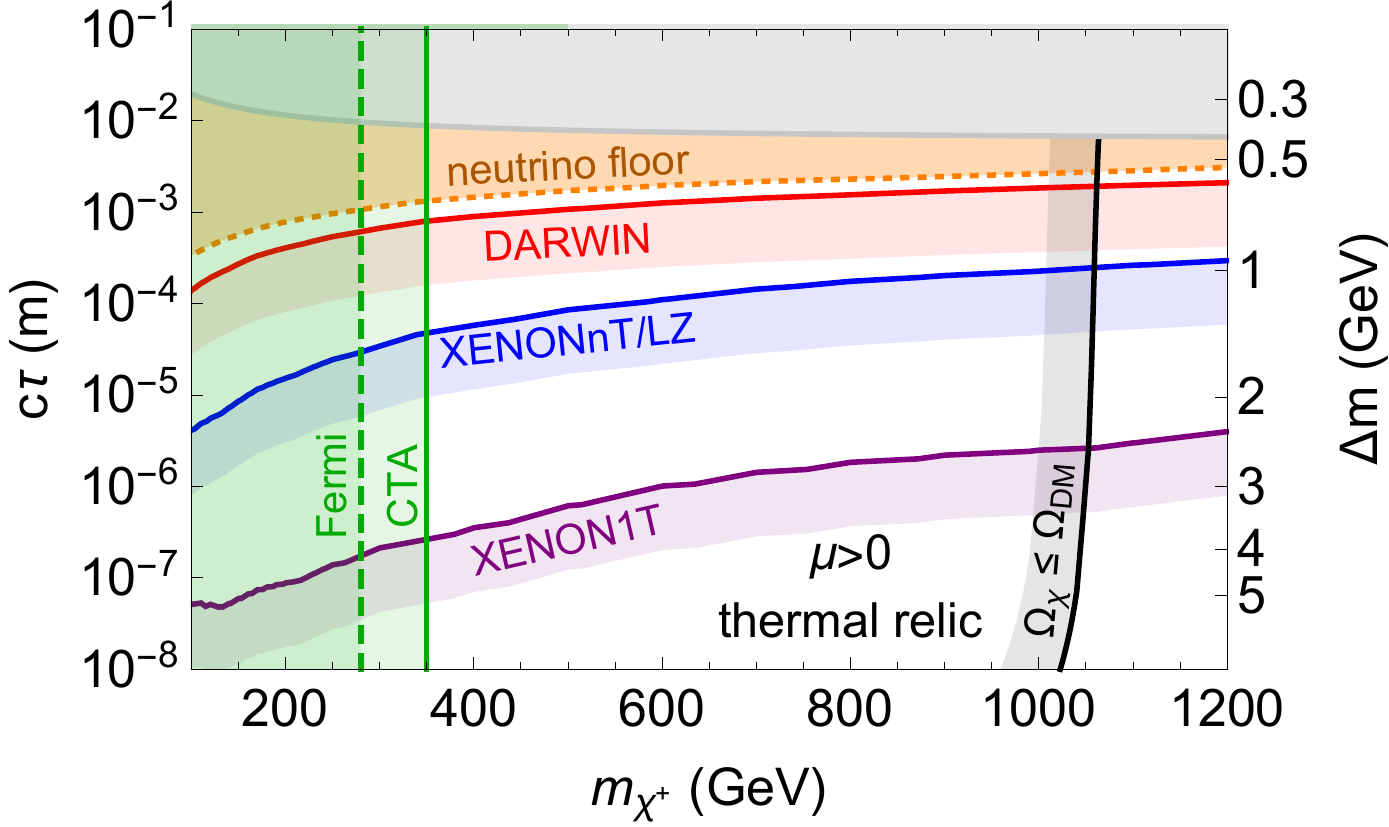}
\end{center}
\caption{
Projected Higgsino bounds from future $pp$ colliders (top) and cosmology (bottom). %
\emph{Top:} 
Vertical bands indicate the approximate projected mass reach of monojet searches, with darker shading indicating the dependence of reach on the assumed  systematic error. 
Regions above black contours can be excluded by disappearing track searches~\cite{Mahbubani:2017gjh} at the HL-LHC (optimistic and pessimistic) and FCC-hh. See text for details.
\emph{Bottom}: 
Longer lifetimes indicate smaller direct detection signal, hence the bounds from XENON1T~\cite{Aprile:2015uzo}, XENONnT~\cite{Aprile:2015uzo}/LZ~\cite{Akerib:2015cja} and DARWIN~\cite{Aalbers:2016jon} are sensitive to the region \emph{below} the colored contours. 
The orange region lies below the neutrino floor for direct detection. 
Also shown is the approximate mass exclusion of Fermi (existing) and CTA (projected).
The black line indicates the maximum mass for the Higgsinos such that their relic abundance is at most $\Omega_{\rm DM}$.
The $\mu < 0$ case is nearly identical.
Relic density and direct detection bounds are taken from~\cite{Mahbubani:2017tba}.
Grey upper region indicates lifetimes corresponding to smaller mass splittings than the minimal electroweak contribution.
}
\label{f.higgsinobounds}
\end{figure}

To understand the unique role $e^- p$ colliders could play in the exploration of  Higgsino parameter space, 
we  briefly review the reach of future $pp$ colldiers, as well as projected cosmological bounds from dark matter direct and indirect detection. 
This is summarized in \fref{higgsinobounds}.

\subsubsection*{Searches at future $pp$ colliders}
The dominant production mode for EWinos at $pp$ colliders are $s$-channel Drell-Yan-like processes. 
The cross section is much larger than at $e^-p$ colliders, which offers opportunities to search for pure Winos with large decay lengths.
A challenge in the high-energy environment of $pp$ collisions is that the SM final state from the chargino decays are often very soft (sometimes just a single pion) which cannot be reliably reconstructed. 
It is therefore difficult to find the corresponding displaced secondary vertex in this environment: the signal gets swamped by the surrounding hadronic activity, and becomes part of the ``hadronic noise''.

One promising search strategy is the so-called ``disappearing track search'', which targets the traces that the long-lived chargino leaves in the tracker of the detector.
This strategy relies on the chargino to reach the first few inner tracking layers, which severely limits the sensitivity for short lifetimes. 
At the HL-LHC the disappearing track searches have a 
mass reach up to $\sim 200 \gev$
with standard tracking if $c \tau \sim 7$mm ($\Delta m = \Delta_\mathrm{1-loop}$)~\cite{Low:2014cba,Mahbubani:2017gjh,Fukuda:2017jmk}. 
Hypothetical upgrades to the HL-LHC trackers in the high-rapidity region could increase mass reach to about 380 GeV.
We show these two scenarios in \fref{higgsinobounds} (top), using the results from \cite{Mahbubani:2017gjh}. 
(This study examined Higgsinos heavier than 200 GeV, but the proposed search would have sensitivity to lower masses as well.)
The pessimistic HL-LHC disappearing track reach projection assumes that the Higgsino must reach a transverse distance of 30cm, while the optimistic projection only requires 10cm. The realistic reach likely lies between these estimates, but we point out that recent ATLAS tracker upgrades should allow for the reconstruction of Higgsinos that travel 12 cm~\cite{Aaboud:2017mpt}. 

At future 100 TeV colliders like the FCC-hh or the SppC with $3 \iab$ of luminosity,\footnote{Since many recent benchmarks assume $30 \iab$ luminosity for future 100 TeV colliders~\cite{Golling:2016gvc,Contino:2016spe}, these reach estimates may be conservative.} disappearing track searches can probe $m_\chi \sim 1.1 \tev$ if $\Delta m \sim \Delta_\mathrm{1-loop}$ assuming a chargino traveling 10cm can be reconstructed, but the reach disappears for shorter lifetimes~\cite{Mahbubani:2017gjh,Fukuda:2017jmk}.\footnote{The reach can be improved considering improved forward tracking close to the beam pipe compared to current benchmark detector proposals.} 
 These sensitivity projections are also shown in \fref{higgsinobounds} (top).

Another strategy is the search for the missing mass that is carried away by the neutral heavy final state. Studies show that such so-called ``monojet searches'' can probe pure Higgsinos with masses up to $\sim 100-200 \gev$ at the HL-LHC \cite{Giudice:2010wb,Schwaller:2013baa,Barducci:2015ffa,Low:2014cba}, depending on assumptions about systematic errors.
At future 100 TeV collider (see e.g. refs.~\cite{Gori:2014oua,Bramante:2014tba,Low:2014cba,diCortona:2014yua,Bramante:2015una}), significantly higher masses of  $\sim 600 - 900 \gev$ \cite{Low:2014cba} can be probed for the {loop-induced} mass splitting. 
We show bounds from~\cite{Low:2014cba} in \fref{higgsinobounds} (top). The darker shading indicates how the mass reach changes when background systematic errors are varied between 1\% and 2\%.\footnote{For larger mass splittings, a soft lepton search can increase Higgsino mass reach \cite{Low:2014cba}, but $\Delta m < 5 \gev$ in our region of interest.}

In general, the direct detection of the chargino LLP yields more information than a monojet missing energy signal.   
Both of the above search strategies suffer significant limitations. Monojet (or mono-X) searches have modest mass reach and reveal no information as to the nature of the produced BSM state beyond the invisibility of the new final states.\footnote{The prospects of the mono-Z searches at the FCC are currently under investigation~\cite{Mahbubani:2017tba}.} It would therefore be impossible to diagnose the signal as coming from a Higgsino-like state. 
Disappearing track searches can have slightly higher mass reach, but only if the lifetime is near the theoretically motivated maximum for this scenario. 

Lifetimes below a few mm are in general extremely challenging to probe in these environments.
It is clear, that the pure Higgsinos with their extremely small mass splitting and relatively short decay length are something of a night-mare scenario for searches at proton-proton colliders.

\subsubsection*{Cosmology}

EWinos make natural candidates for thermal Dark Matter if they are stable on cosmological time scales.
Thus, cosmological considerations may serve as general motivator for our theoretical setup and provide constraints for specific models.
It is important to keep in mind, however, that these constraints are dependent on the universe's cosmological history, and are therefore not as robust as collider searches.

Assuming that the 
lightest neutralino contributes to the thermal relic density provides us with additional bounds from cosmological observation. 
The abundance from Higgsinos with masses above $\sim 1.1 \tev$~\cite{ArkaniHamed:2006mb} is larger than the observed dark matter relic density. 
This makes 1.1 TeV an obvious target for collider searches, see \fref{higgsinobounds} (bottom)

Direct dark matter detection experiments are sensitive to Higgsinos with mass splittings in the GeV range or above, see e.g. ref.~\cite{Barducci:2015ffa}.
Sensitivity projections are summarized in \fref{higgsinobounds} (bottom), and notably constrain short lifetimes but not long ones.
This is due to the coupling to the Higgs boson, which mediates nuclear scattering and depends on the Higgsino-Bino mixing angle, or, equivalently, $\delta m - \Delta_\mathrm{1-loop}$ and only becomes appreciable for mass splittings $\sim \gev$. Hence, the lack of signals in direct detection strongly favors a highly compressed spectra.\footnote{It is also possible to have an accidentally small (or null) coupling of Higgs to dark matter in the so called blind-spots~\cite{Cheung:2012qy}. We will not consider this option further in this work.}
The most sensitive of these future experiments is DARWIN~\cite{Aalbers:2016jon}, which will be able to probe DM-nucleon cross sections very close to the so-called \emph{neutrino floor}, where backgrounds from solar, cosmic and atmospheric neutrinos become relevant. For thermal Higgsino DM, this scattering rate corresponds to mass splittings of about 0.5 GeV.\footnote{This implies a lower bound on the singlet mass of 10 TeV. The singlet might then be well outside the reach of both the present and future generation of collider experiments.} Probing cross sections below the neutrino floor will be much more challenging.
 
Indirect detection experiments search for signs of dark matter annihilation in the cosmic ray spectra.
Assuming a thermal relic abundance, current bounds from Fermi disfavor masses below 280 GeV, with proposed CTA measurements being sensitive to $m_\chi \sim 350 \gev$ \cite{Calibbi:2015nha}. AMS antiproton data might exclude somewhat higher masses \cite{Krall:2017xij}, but that bound is subject to very large uncertainties.

While these cosmological bounds complement collider searches, they are much more model-dependent. One can imagine a Higgsino-like inert doublet scenario which does not give rise to a stable dark matter candidate (e.g. the lightest neutral state could decay to additional hidden sector states), making colliders the only direct way to probe their existence.
Even if the assumptions about cosmology hold, collider searches are vital to fill in the blind spots below the neutrino floor. If a direct detection signal is found, the precise nature of dark matter would then have to be confirmed with collider searches. 
Finally, even with the most optimistic projections there are regions of parameter space at intermediate mass splitting (lifetimes $\lesssim$ mm) that are difficult to probe using both direct detection and current strategies at $pp$ colliders. 

\subsection{Higgsino search at $e^- p$ colliders}

\begin{figure}[t]
\begin{center}
\includegraphics[width=4.2cm]{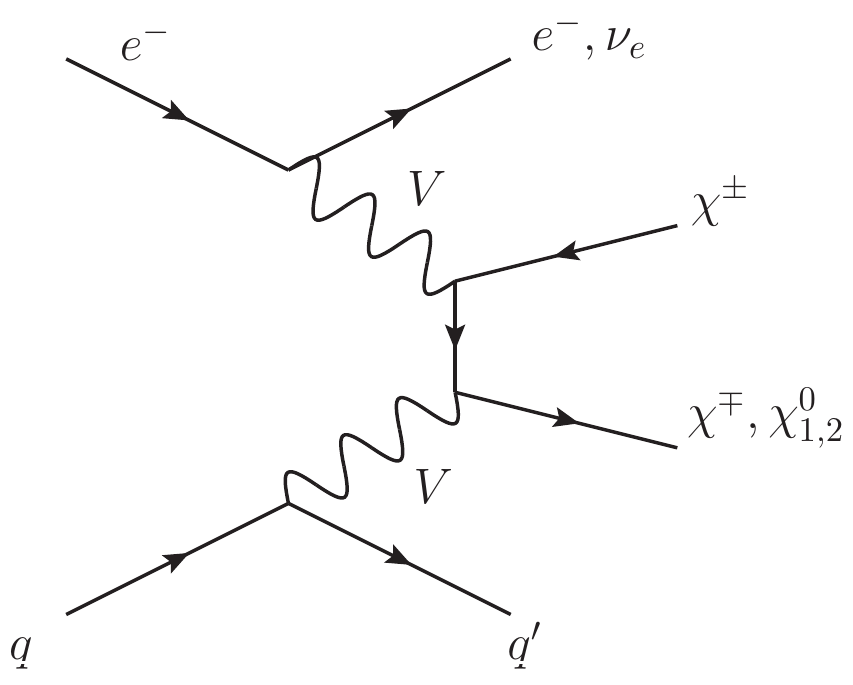}
\includegraphics[width=4.2cm]{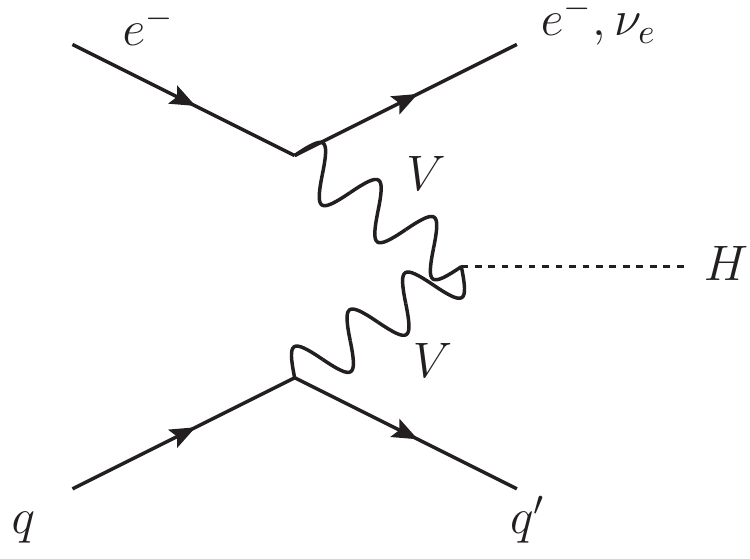}
\end{center}
\caption{
Example of dominant Higgsino (left) and Higgs (right) production processes at $e^- p$ colliders. $V = W^\pm$ or $Z$ as required.
}
\label{f.feynman}
\end{figure}

\begin{figure}[t]
\begin{center}
\includegraphics[width=8cm]{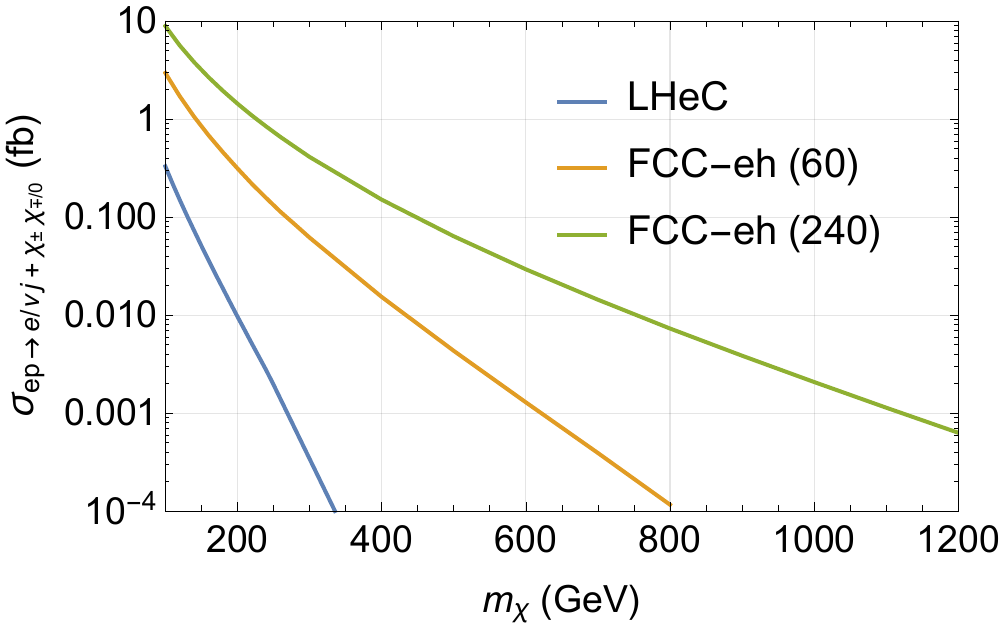}
\end{center}
\caption{
Production rate of Higgsinos at $e^- p$ colliders. The fraction of events with two charged Higgsino LLPs is $\sim 40-50\%$. }
\label{f.productionxsec}
\end{figure}

At $e^- p$ colliders, Higgsinos are produced dominantly in VBF processes as shown in \fref{feynman} (left). Since the production process is $2 \to 4$ it suffers significant phase space suppression and has a rather small cross section, as shown in \fref{productionxsec}. Fortunately, the spectacular nature of the LLP signal, and the clean experimental environment, still allows for significant improvements in reach compared to the existing search strategies outlined in the previous subsection.

\subsubsection*{LLP signature}
We first consider searches at the LHeC. 
Weak-scale Higgsinos are produced in association with a recoiling, highly energetic jet with $p_T > 20 \gev$. This jet alone will ensure that the event passes trigger thresholds and is recorded for offline analysis. 
Crucially, the measurement of this jet will also determine the position of the primary vertex (PV) associated with the Higgsino production process.

Due to the asymmetric beams the center-of-mass frame of the process is boosted by $b_\mathrm{com} \approx \frac{1}{2}\sqrt{{E_e}/{E_p}} \approx 5.5$ with respect to the lab frame. 
Subsequently, the long lived charginos are typically significantly boosted along the proton beam direction, which increases their lifetime in the laboratory frame.

For small mass splittings $\lesssim 1 \gev$ considered here, the dominant decay modes of the Higgsinos are to single $\pi^\pm, e^\pm, \mu^\pm$ + invisible particles. 
The single visible charged particle typically has transverse momenta in the $\mathcal{O}(0.1 \gev)$ range. 
In the clean environment (i.e.\ low pile up) of the $e^- p$ collider, such single low-energy charged tracks can be reliably reconstructed.

\subsubsection*{Analysis strategy}

The following offline analysis strategy is sketched out in \fref{displaced}. 
One or two charginos are produced at the PV, which is identified by the triggering jet (A). 
A chargino decaying to a single charged particle is depicted in \fref{displaced} (B). 
The charged track has an impact parameter with respect to the PV. 
If the impact parameter with respect to the PV is greater than a given $r_{min}$ we can tag this track as originating from an LLP decay, which holds also when the LLP decays within the interaction region.
This heavily relies on backgrounds due to pile-up being either absent or controllable.

If the chargino decays to two or more charged particles, a conventional displaced vertex can be reconstructed (C). In that case, the PV-DV distance has to be greater than $r_{min}$ to identify an LLP decay.\footnote{In a realistic analysis, $r_{min}$ can be different for displaced tracks and vertices, but for our analysis it is sufficient to take them to be identical.}

The most relevant parameter of our search strategy is thus $r_\mathrm{min}$.
Our benchmark value is $r_{min} = 40 \mu m$, which corresponds to about 5 nominal detector resolutions. We also consider the case of 5 `optimistic' detector resolutions  ($r_{min} = 25 \mu m$) and a pessimistic scenario with $r_{min} = 80 \mu m$.
Moreover, the $p_T$ threshold for reconstruction of a single charged particle is also relevant. In order to study the impact of the $p_T$ threshold, we will consider a benchmark value of $p_T^\mathrm{min} = 100 \mev$, corresponding to a gyromagnetic radius of $\mathcal{O}(10 \mathrm{cm})$ {for the B field of $3.5$ T}. We also consider an optimistic scenario of $p_T^\mathrm{min} = 50 \mev$ and a pessimistic scenario of $p_T^\mathrm{min} = 400 \mev$, which corresponds to the threshold for track ID at ATLAS and CMS in a high pile-up environment~\cite{atlasidtdr}. \footnote{At an $e^-p$ collider the full four momentum can be measured, and employing $|p|$ rather than $p_T$ would lead to a slight increase in sensitivity.
However, in order to be comparable with $pp$ collider thresholds, we use $p_T$ in the following.}

We assume 100\% reconstruction efficiency for displaced tracks and vertices. 
The estimation of the realistic (expected-to-be $\mathcal{O}(1)$) efficiencies requires a full simulation of the detector response to our signal, which is beyond the scope of our paper and will be left for future work. We do not expect this to significantly affect our conclusions.

\begin{figure}[t]
\begin{center}
\includegraphics[width=8cm]{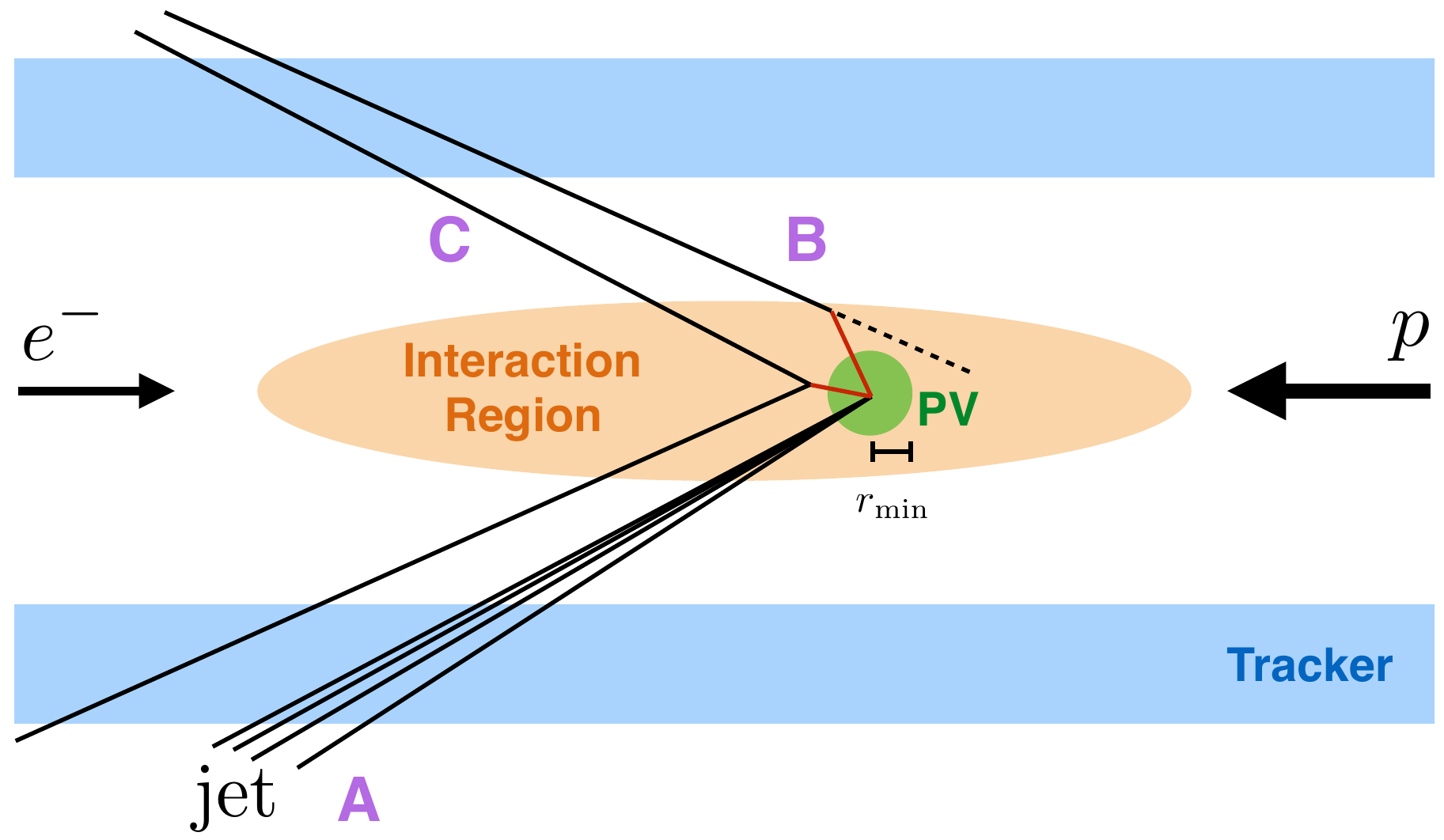}
\end{center}
\caption{
Sketch of our LLP search strategy at $e^- p$ colliders. 
Single or pair-production of weak-scale Higgsino LLPs (red)  is practically always associated with the production of a hard jet (A) with $p_T > 20 \gev$ and $|\eta| < 4.7$ which reaches the tracker and passes the trigger. The charged jet constituents (black) identify the primary vertex (PV).
For Higgsinos decaying into $e/\mu/\pi^\pm + \chi^0_{1,2}$ (B), the LLP is detected if the charged particle trajectory (black solid and dashed) is reconstructed with $p_T > p_T^\mathrm{min}$ and has impact parameter greater than $r_{min}$. 
For LLPs decaying into two or more charged particles (C), a DV can be reconstructed, and the LLP is identified if the distance to the PV is more than $r_{min}$.
The electron or neutrino in the event as well as neutral final states of LLP decay are not shown.  
}
\label{f.displaced}
\end{figure}

\subsubsection*{Event simulation and analysis}

The production of MSSM Higgsinos is simulated in \texttt{MG5\_aMC@NLO}~\cite{Alwall:2014hca} at parton-level, which is sufficient given the almost purely geometrical nature of our signal. 
For each  chargino $k$ the probability of detecting it as an LLP is 
\begin{equation}
P^{(k)}_\mathrm{detect} = \sum_i \mathrm{Br}_i(\Delta m(c \tau)) P_i
(c \tau) \ ,
\end{equation}
where $k = 1, 2$ for chargino pair production events.
The index $i$ stands for the decay processes in \fref{higgsinodecays}, with branching ratios $\mathrm{Br}_i$. 
$P_i$ is the probability of detecting this particular chargino if it decays via process $i$. For 2- and 3-body decays to a single charged particle, it is computed by choosing the charged particle momentum from the appropriate phase space distribution in the chargino rest frame, then computing the minimum distance the chargino must travel for the impact parameter of the resulting charged track to be greater than $r_{min}$. $P_i$ is the chance of the chargino traveling at least that distance given its boost and the chosen lifetime $c \tau$. $P_i = 0$ if the charged particle $p_T$ lies below threshold or it does not hit the tracker. 

For decays to ``jets'', defined as three charged pions (all hadronic decays) for $\Delta m$ below (above) $\Delta m_*$, we examine two possibilities. Optimistically, one would expect the jet to contain two or more relatively energetic charged particles, allowing a DV to be reconstructed. $P_{jet}$ is then computed simply by requiring the chargino to travel at least $r_{min}$ from the PV. Pessimistically the jet has to contain at least one charged particle, and we assign $P_{jet} = P_{\pi^\pm \pi^0 \pi^0}$. The difference between the optimistic and pessimistic $P_{jet}$ scenarios represents an uncertainty on our sensitivity estimate. 

For each event with one chargino, $P^{(1)}_\mathrm{detect}$ represents the chance of detecting a single LLP in the event. For each event with two charginos, $1 - (1-P^{(1)}_\mathrm{detect})(1-P^{(2)}_\mathrm{detect})$ is the chance of observing at least one LLP, while $P^{(1)}_\mathrm{detect} P^{(2)}_\mathrm{detect}$ is the chance of observing two LLPs. This allows us to compute the number of observed events with at least one or two LLPs, $N_\mathrm{1+LLP}$ and $N_\mathrm{2LLP}$, as a function of chargino mass and chargino lifetime. 

We show contours of $N_\mathrm{1+LLP}$ and $N_\mathrm{2LLP}$ in \fref{higgsinoLHeC} for $\mu > 0$. 
The darker (lighter) shading represents the contour with the lowest (highest) estimate of event yield, obtained by minimizing (maximizing) with respect to the two hadronization scenarios of $m_d = 0$ or $0.5 \gev$, and adopting the pessimistic (optimistic) $P_{jet}$ reconstruction assumption. The difference between the light and dark shaded regions can be interpreted as a range of uncertainty in projected reach.\footnote{We note that the abrupt ``bite'' in the green shaded region of the top plot around $(m_\chi, c\tau) \sim (140 \gev, 10^{-5} \mathrm{m})$ is an artifact of assuming 100\% DV reconstruction once the Higgsino decays to jets of two or more charged particles turn on at larger mass splitting (under the optimistic reconstruction assumption). In reality, this intermediate region would likely be smoothly interpolated by a gradual turn-on, when more efficiently reconstructed DVs start dominating over displaced single tracks.} The $\mu < 0$ case is very similar in all of our studies, so we only show the positive case.

\begin{figure}
\begin{center}
\includegraphics[width=8cm]{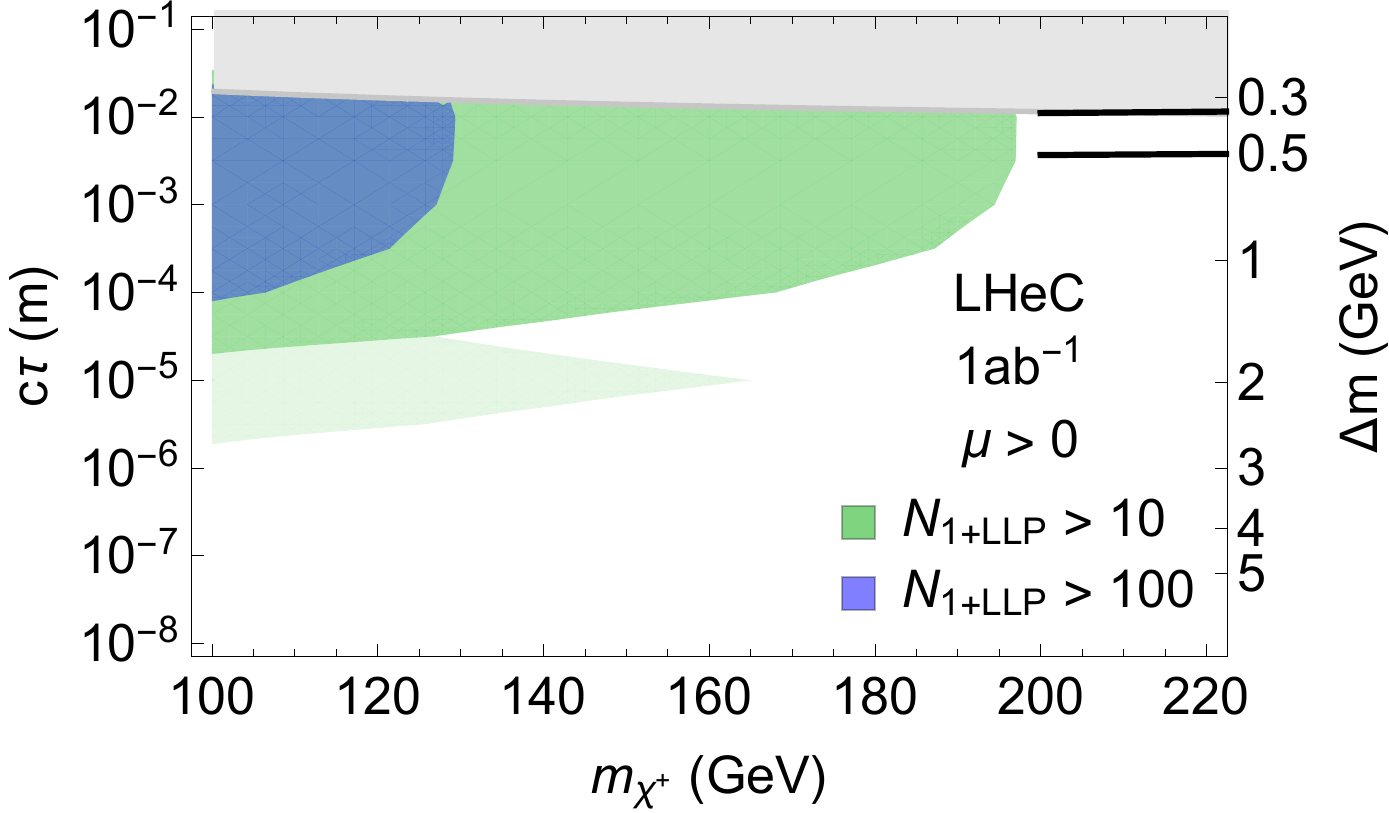}
\\
\includegraphics[width=8cm]{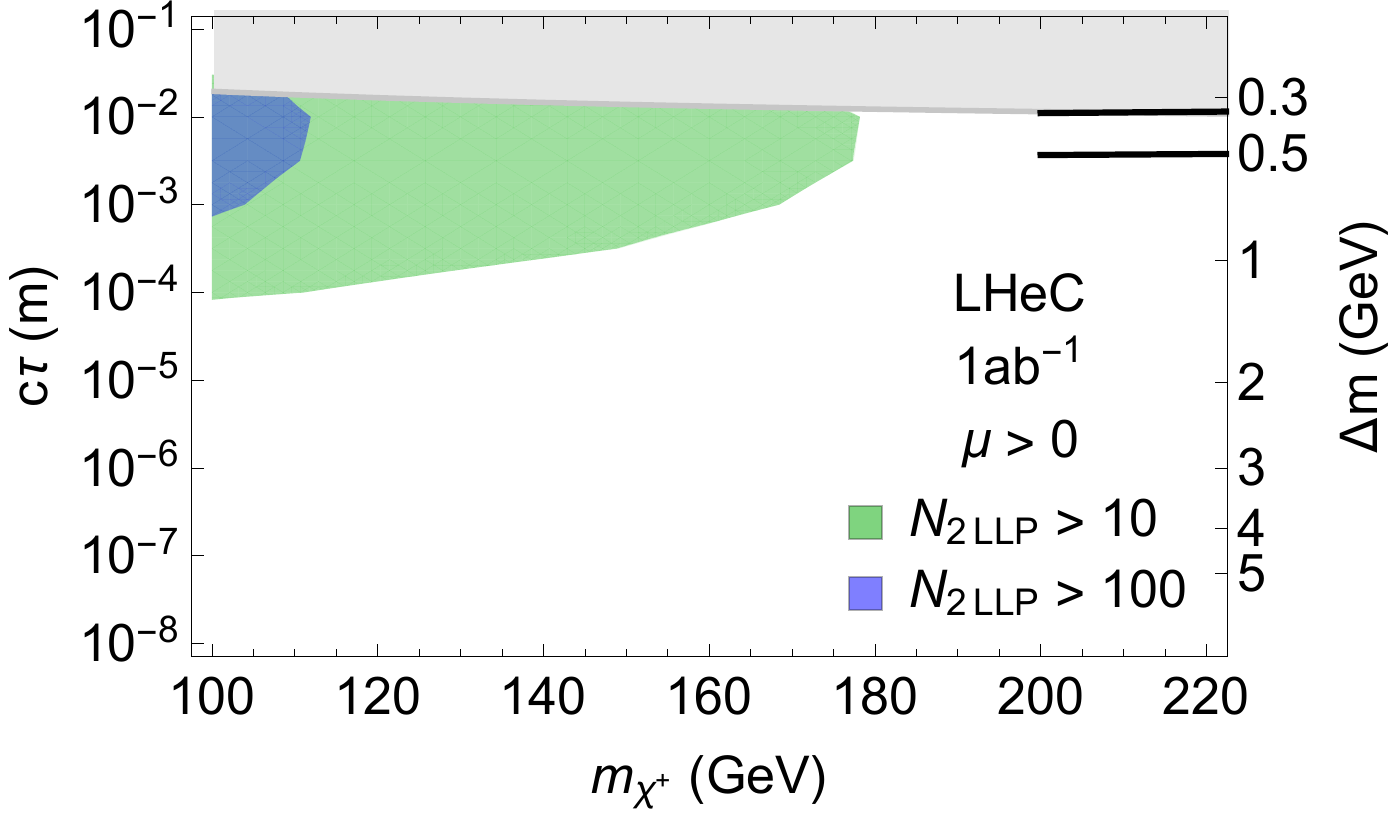}
\end{center}
\caption{
Regions in the $(m_{\chi^\pm}, c \tau)$ Higgsino parameter plane where more than 10 or 100 events with at least one (top) or two (bottom) LLPs are observed at the LHeC. Light shading indicates the uncertainty in the predicted number of events due to different hadronization and LLP reconstruction assumptions. 
Approximately 10 signal events should be descernable against the $\tau$-background at $2\sigma$, in particular for 2 LLPs, so the green shaded region represents an estimate of the exclusion sensitivity. 
For comparison, the black curves are the optimistic and pessimistic projected bounds from HL-LHC disappearing track searches, see \fref{higgsinobounds}.}
\label{f.higgsinoLHeC}
\end{figure}

\begin{figure*}
\begin{center}
\begin{tabular}{cc}
\includegraphics[width=8cm]{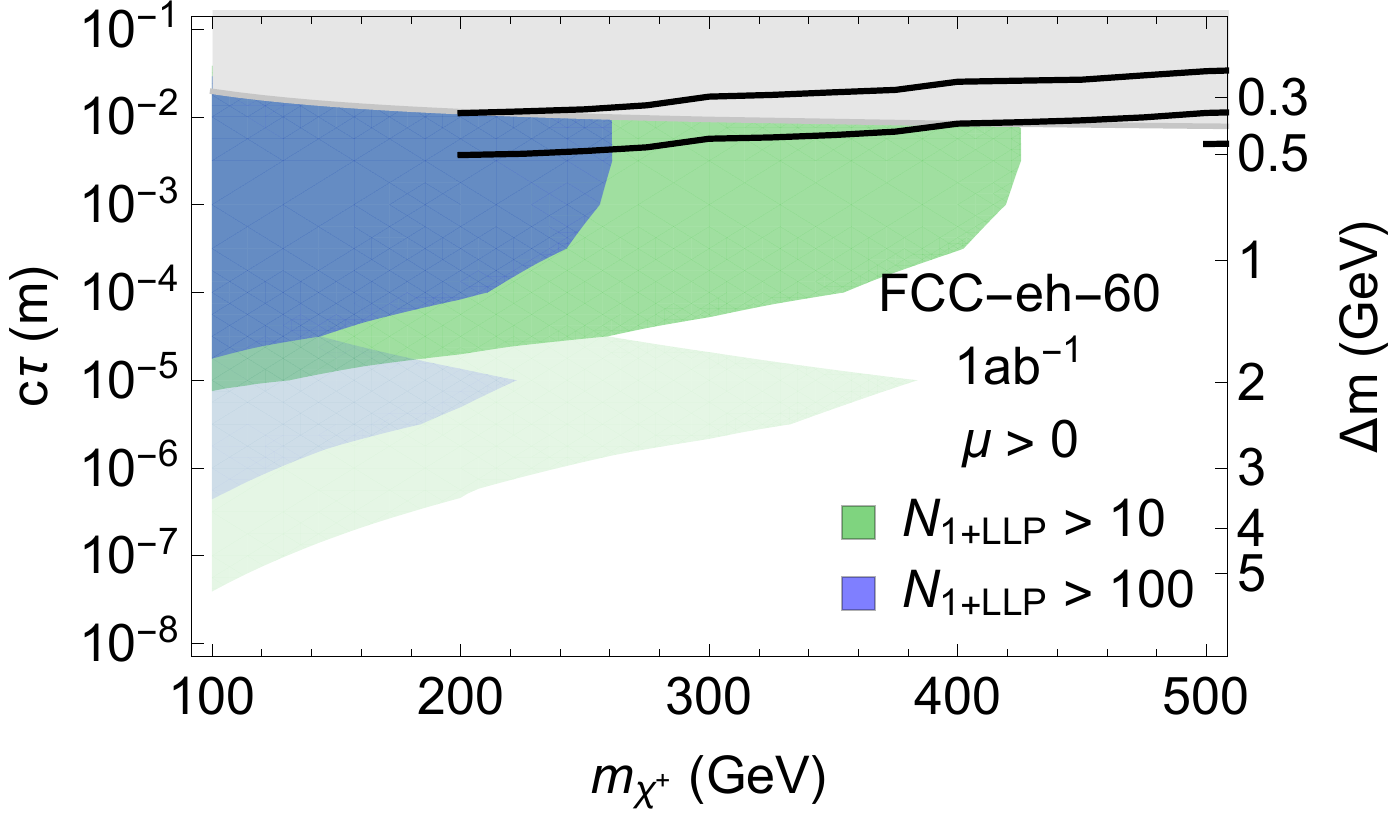}
&
\includegraphics[width=8cm]{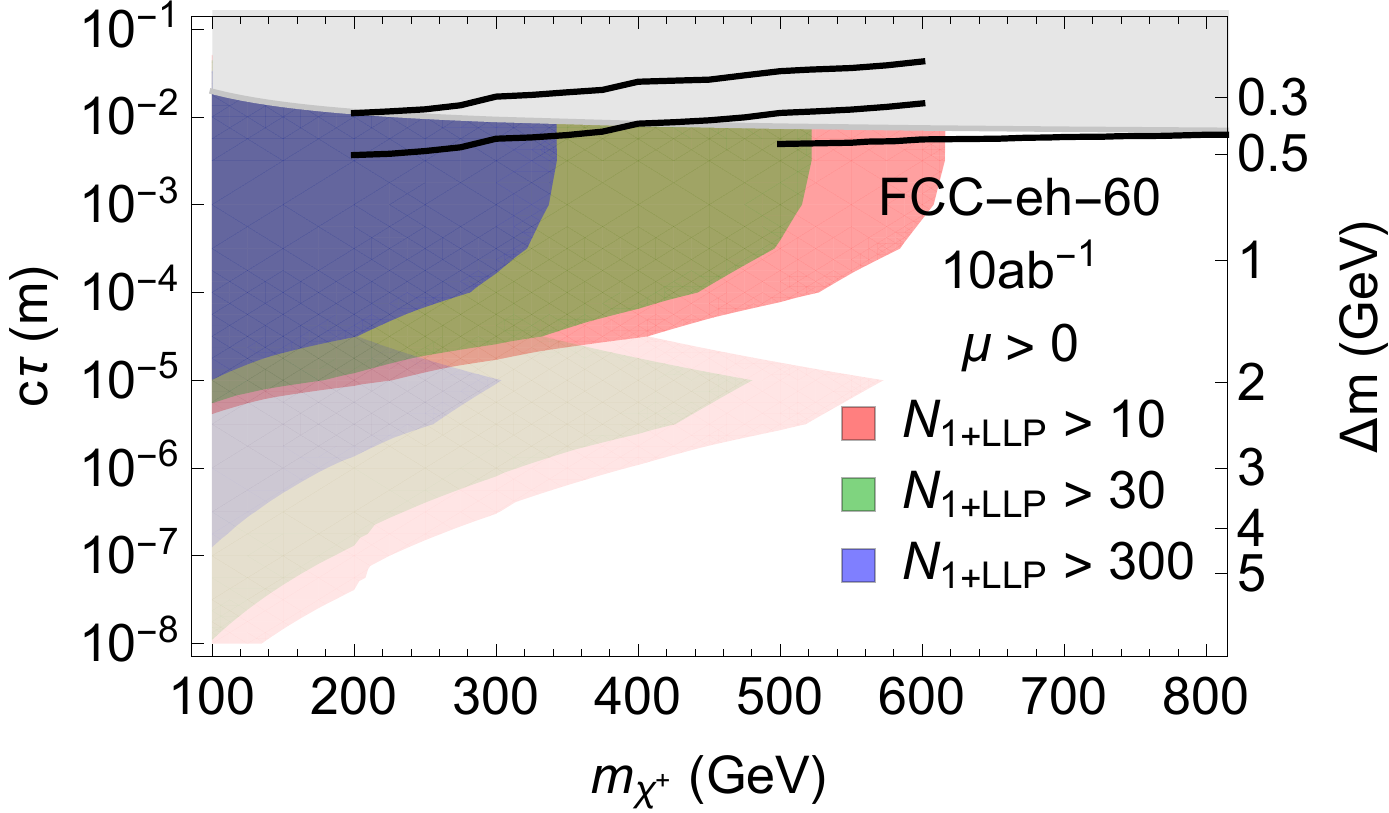}
\\
\includegraphics[width=8cm]{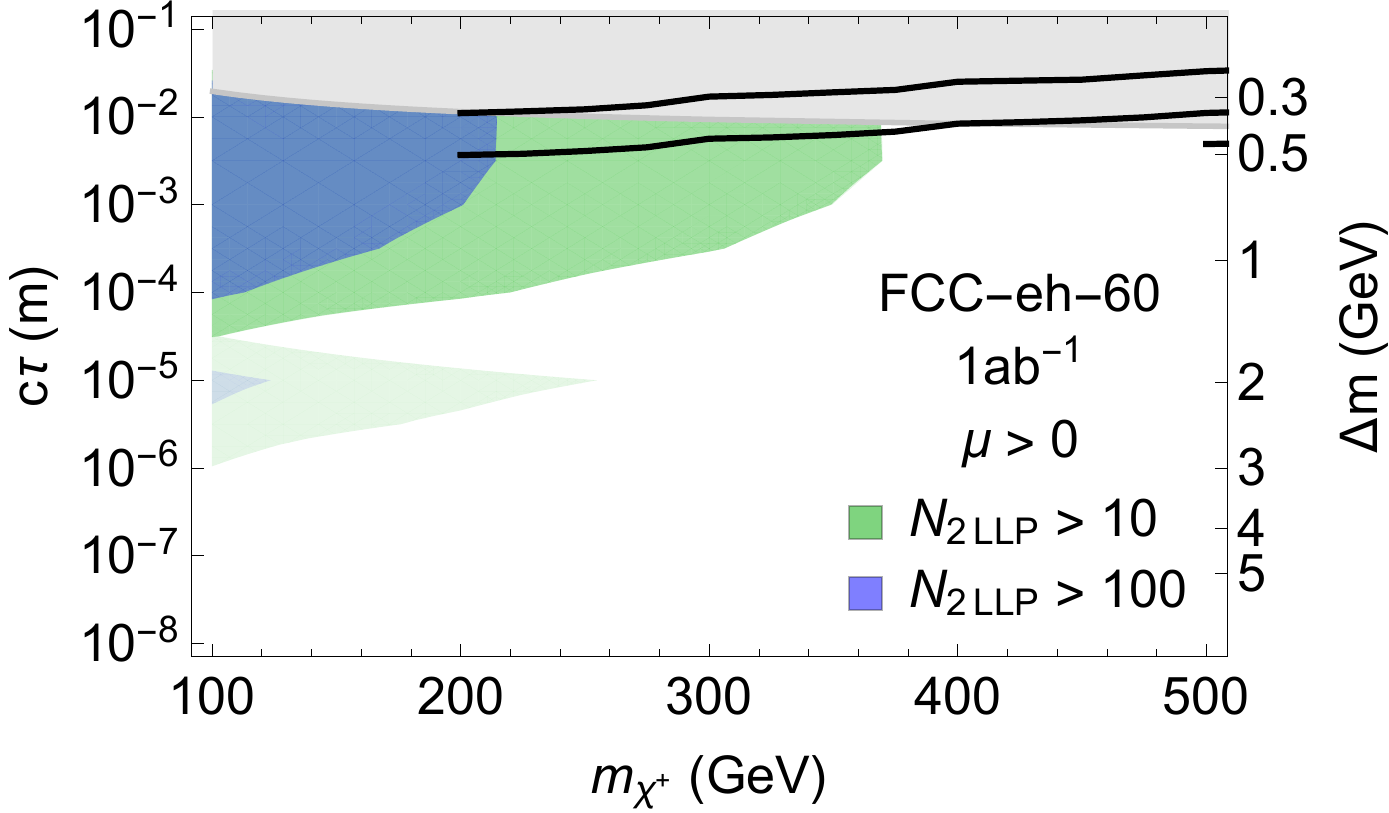}
&
\includegraphics[width=8cm]{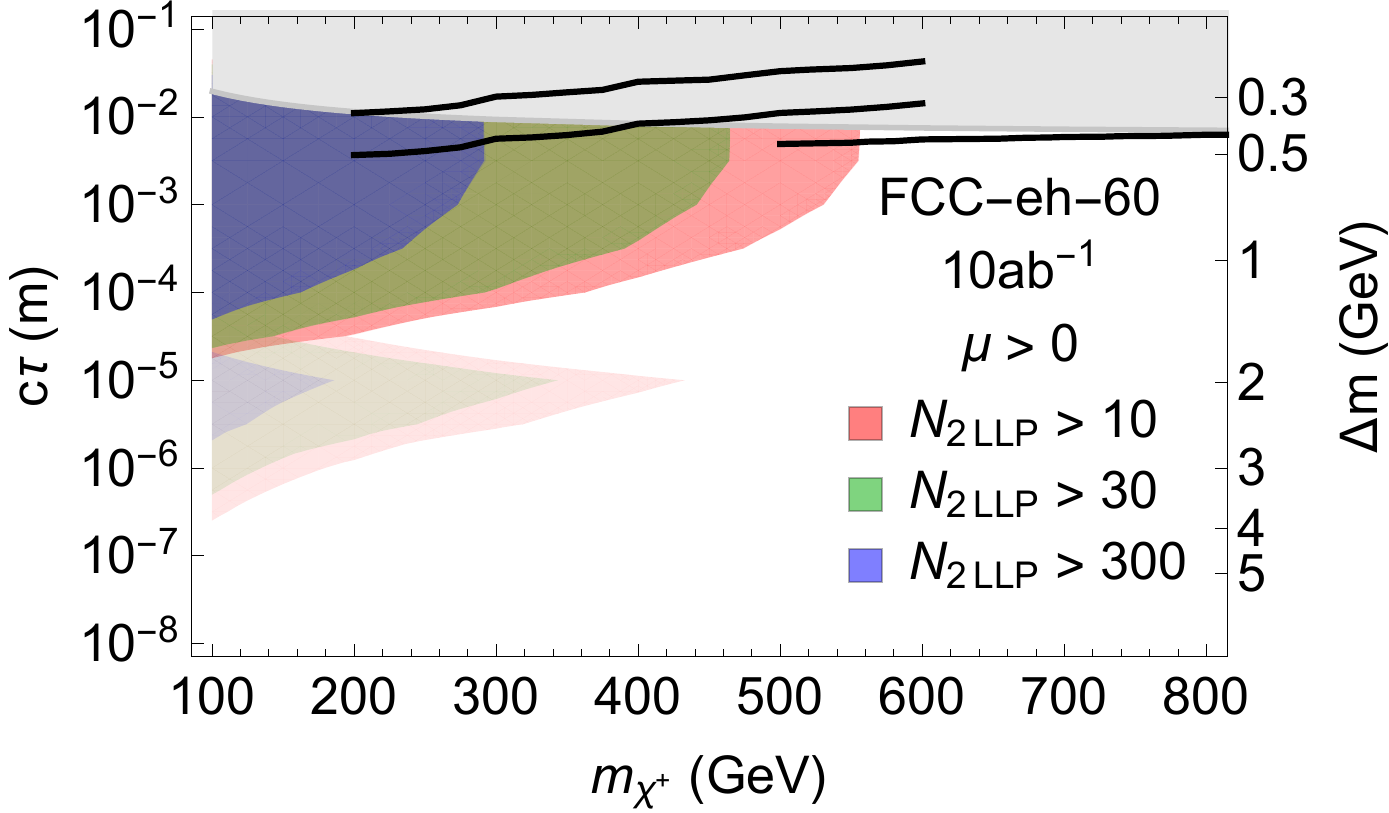}
\end{tabular}
\end{center}
\caption{
Regions in the $(m_{\chi^\pm}, c \tau)$ Higgsino parameter plane where more than the indicated number of one (top) or two (bottom) LLPs are observed at the FCC-eh with a 60 GeV electron beam and 1 $\iab$ (left) or 10 $\iab$ (right) of luminosity. Light shading indicates the uncertainty in the predicted number of events due to different hadronization and LLP reconstruction assumptions. 
As for the LHeC estimate in \fref{higgsinoLHeC},  the green region represents our $2\sigma$ sensitivity estimate in the presence of $\tau$ backgrounds. For 10 $\iab$, red shading is an optimistic sensitivity estimate in case background rejection is better than we anticipate.
For comparison, the black curves are projected bounds from disappearing track searches, for the HL-LHC (optimistic and pessimistic) and the FCC-hh, see \fref{higgsinobounds}.
}
\label{f.higgsinoFCCeh60}
\end{figure*}

\begin{figure*}
\begin{center}
\begin{tabular}{cc}
\includegraphics[width=8cm]{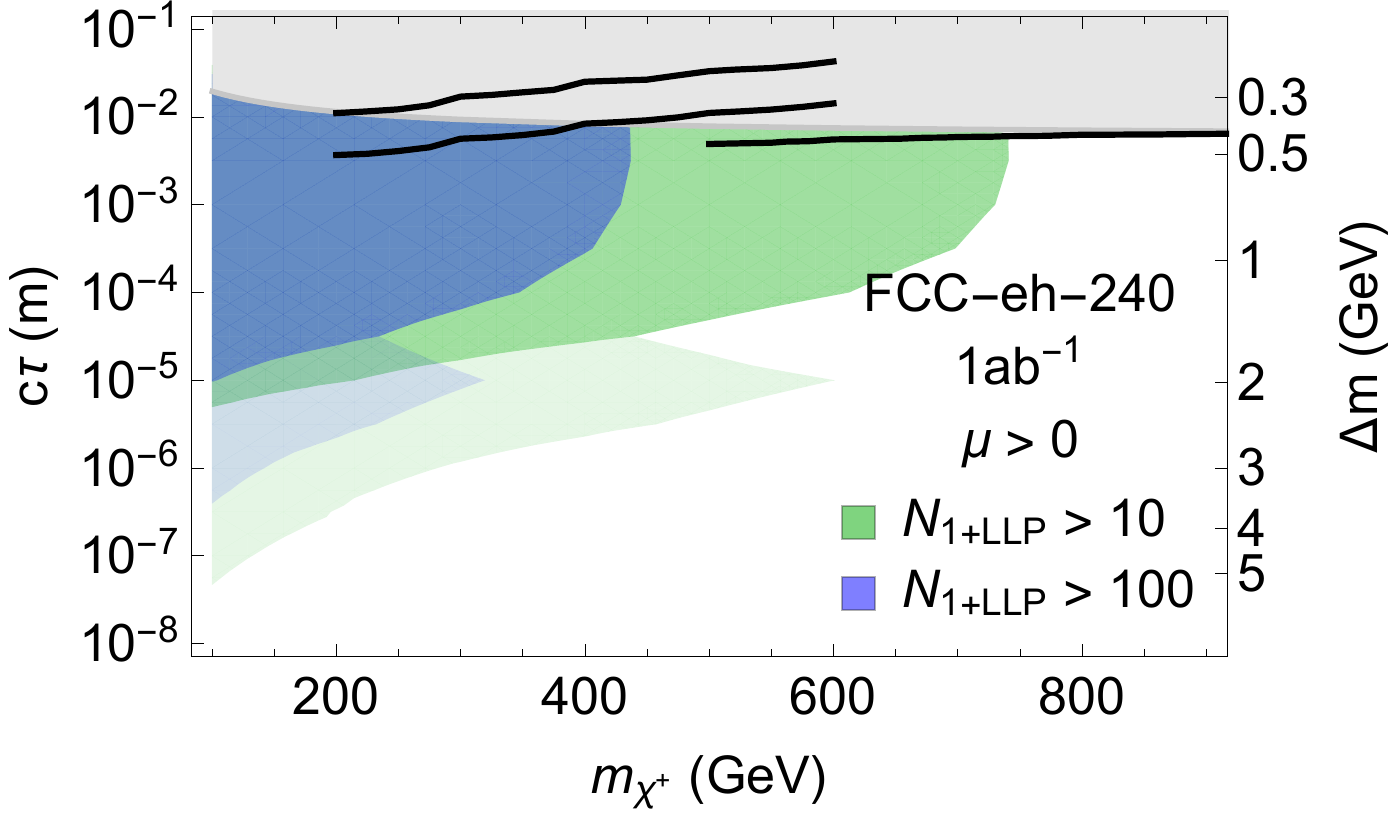}
&
\includegraphics[width=8cm]{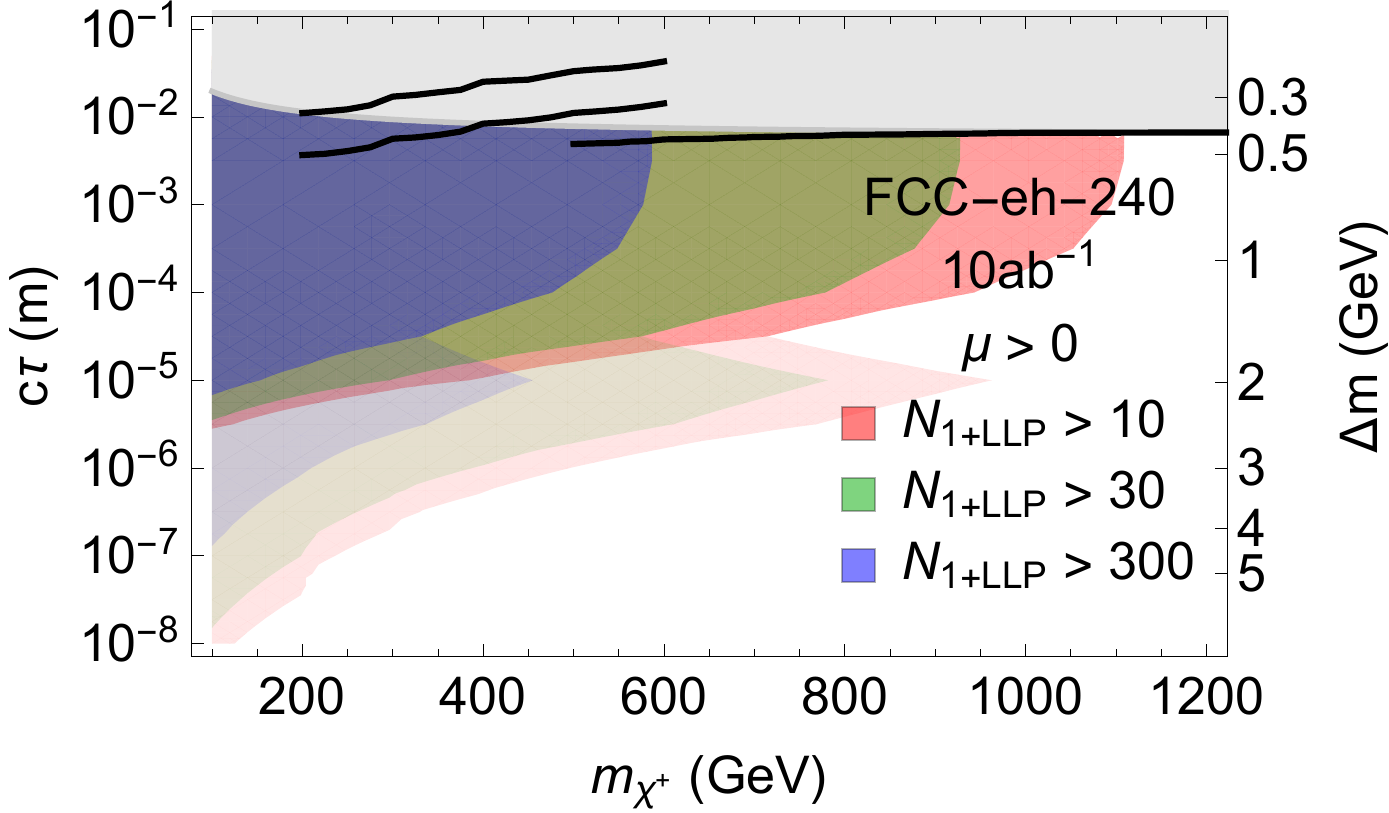}
\\
\includegraphics[width=8cm]{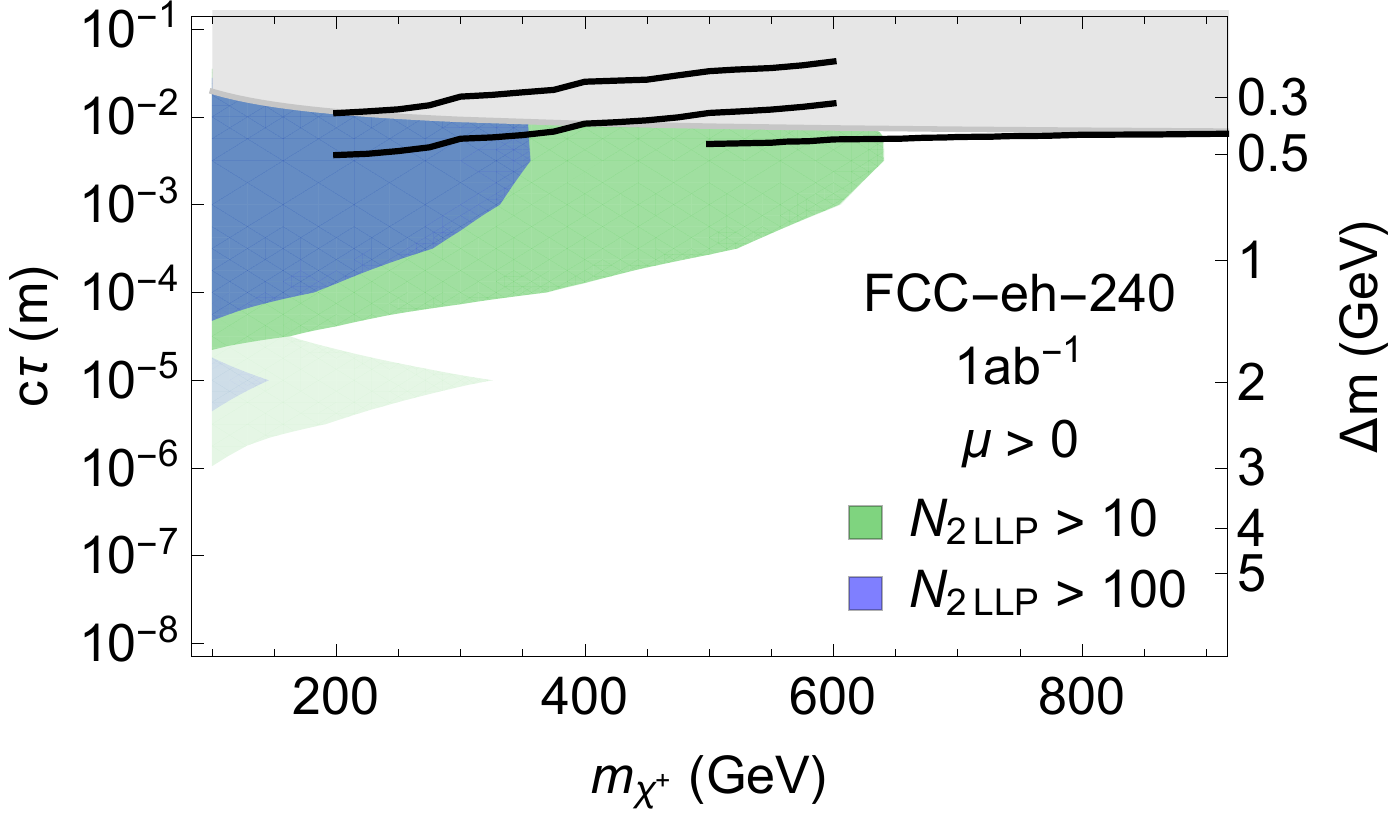}
&
\includegraphics[width=8cm]{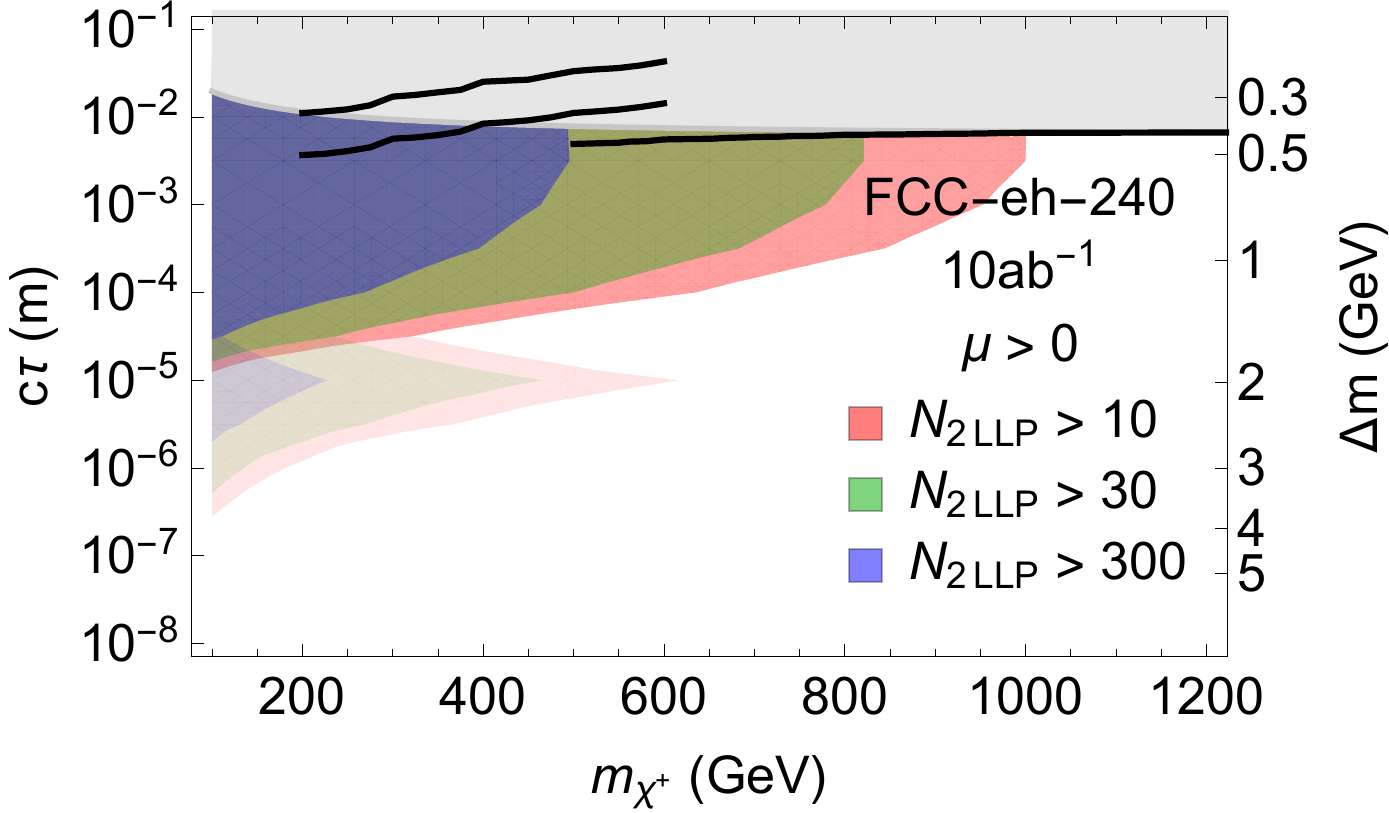}
\end{tabular}
\end{center}
\caption{
Same as \fref{higgsinoFCCeh60} for the FCC-eh with a 240 GeV electron beam.
}
\label{f.higgsinoFCCeh240}
\end{figure*}

\subsubsection*{Backgrounds}

An important and irreducible background SM background to our LLP signature is the decays of tau leptons, which have a proper lifetime of $\sim 0.1$mm and beta-decay into the same range of final states as the charginos.
Events with one ($\tau^+ \nu_\tau$) and two taus ($\tau^+ \tau^-$) are produced via VBF together with a jet with $p_T > 20 \gev,  |\eta| < 4.7$ at LHeC with cross sections of $\sim0.6$ and $\sim0.3$ pb, respectively. 

Since the $\tau$'s originate from the decay of on-shell $W$ and $Z$ bosons, their decay products are much more central and energetic than those of charginos. 
Consequently, despite this background being much larger than the Higgsino signal, it can be suppressed considerably with simple kinematic cuts. 

Specifically, by requiring the final states of LLP decay to be forward ($|\eta| > 1$ in the proton beam direction), the missing energy to be high (MET $\gtrsim 30 \gev$) and the LLP final state energy to be very low ($\lesssim 1.5 \Delta m$ for a given chargino lifetime), a background rejection of $10^{-3}$ ($10^{-4}$) can be achieved for events requiring at least one (two) reconstructed LLPs while keeping a large $\mathcal{O}(1)$ fraction of the Higgsino signal. 

Given the above background cross sections, the number of signal events that would be excludable at the 95\% confidence level ($2\sigma$) above the background are then about 50 (10) for at least one (two) observed LLPs. 
This purely kinematic background rejection is very effective, but still underestimates the sensitivity. In the space of possible final states and decay lengths, $\tau$'s will populate very different regions than the chargino signal. 
While an in-depth study of such an analysis is beyond our scope, a comparison of the observed LLP data to a background template in that space will clearly increase sensitivity even further. 

It is with this in mind that we have shown contours of $N_{1+\mathrm{LLP}, 2\mathrm{LLP}} > 10$ and $>100$. By the above arguments, the former constitutes a realistic expectation for the approximate number of LLPs which should be excludable at 2$\sigma$, while the latter shows how sensitivity is affected if backgrounds are much harder to reject than we anticipated.

\begin{figure*}
\begin{center}
\begin{tabular}{cc}
\includegraphics[width=8cm]{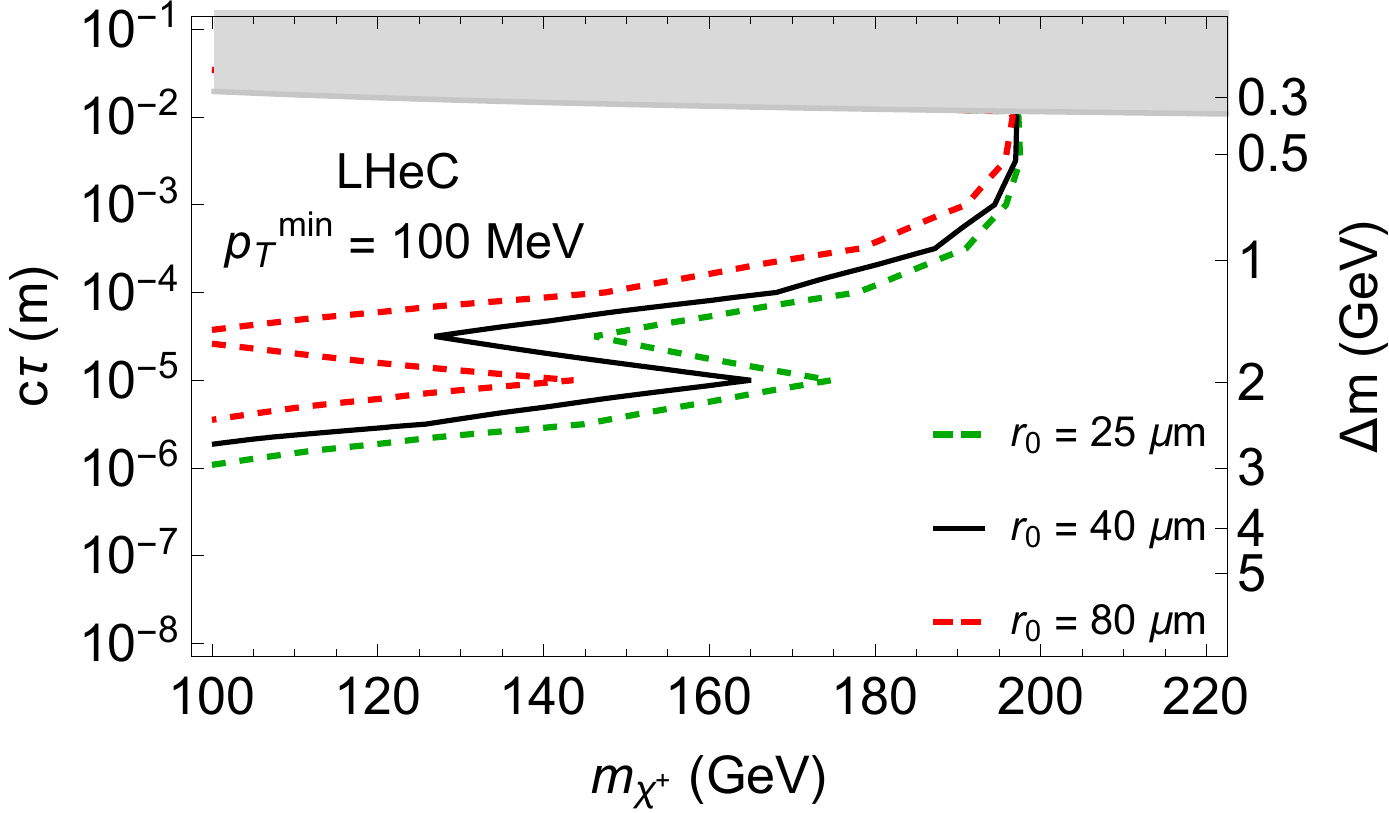}
&
\includegraphics[width=8cm]{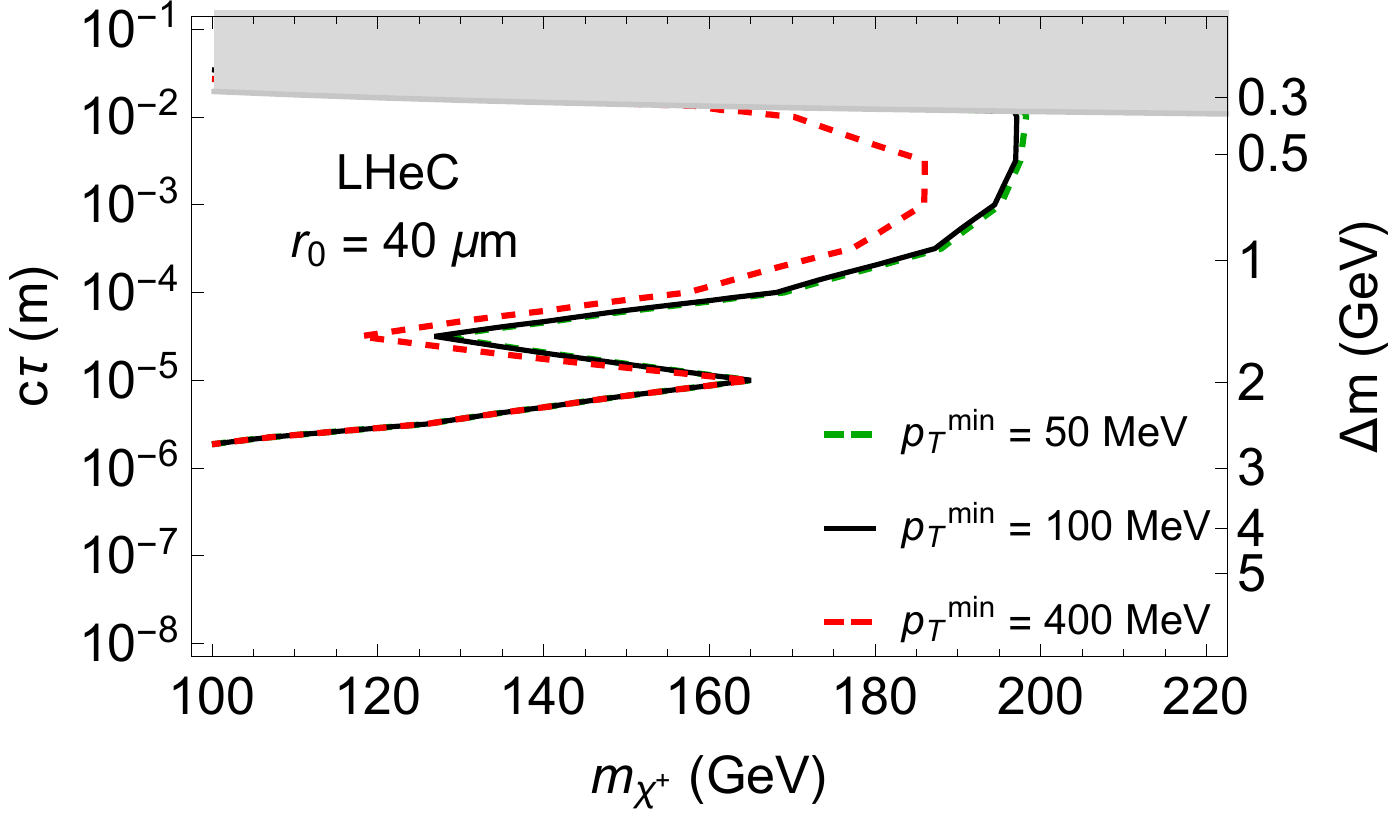}
\\
\includegraphics[width=8cm]{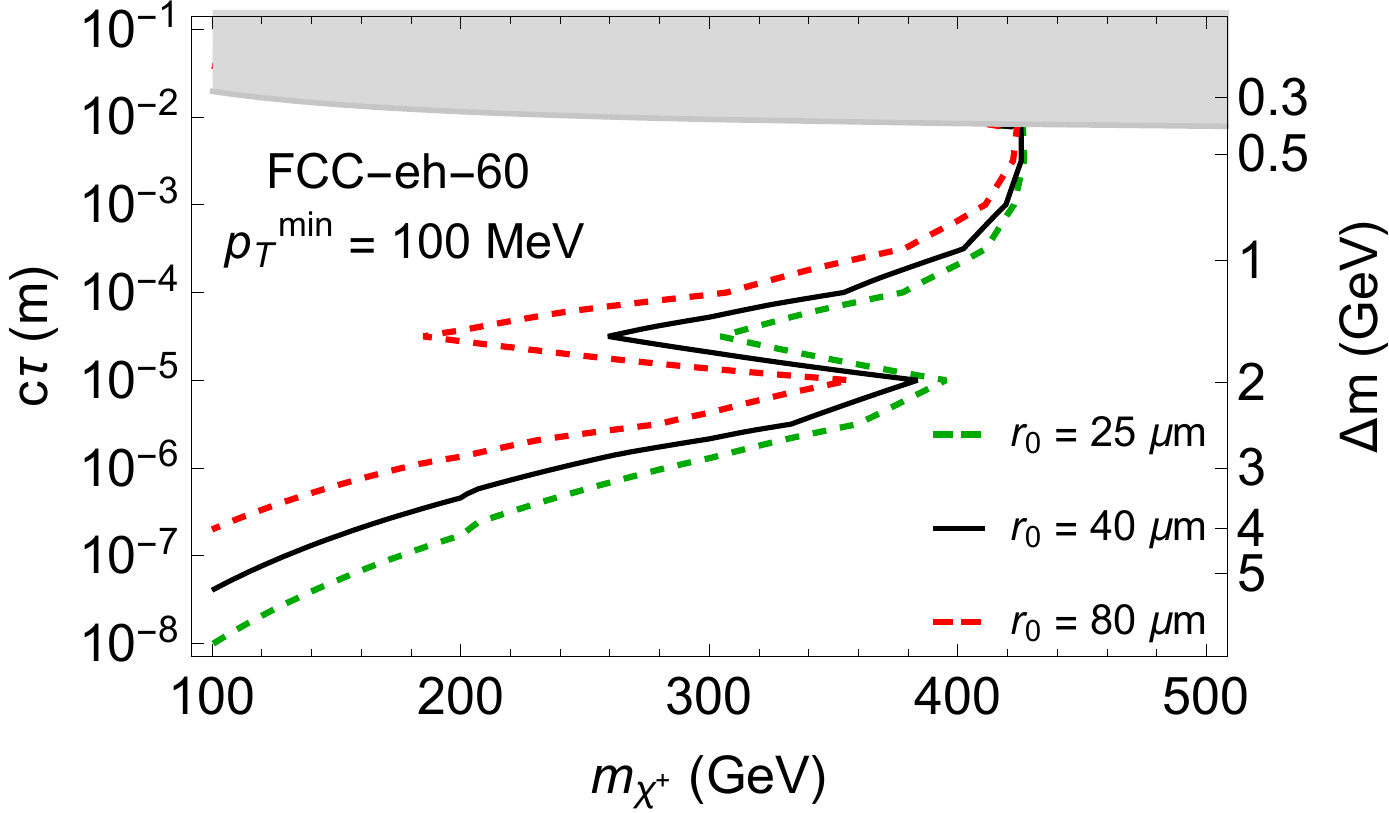}
&
\includegraphics[width=8cm]{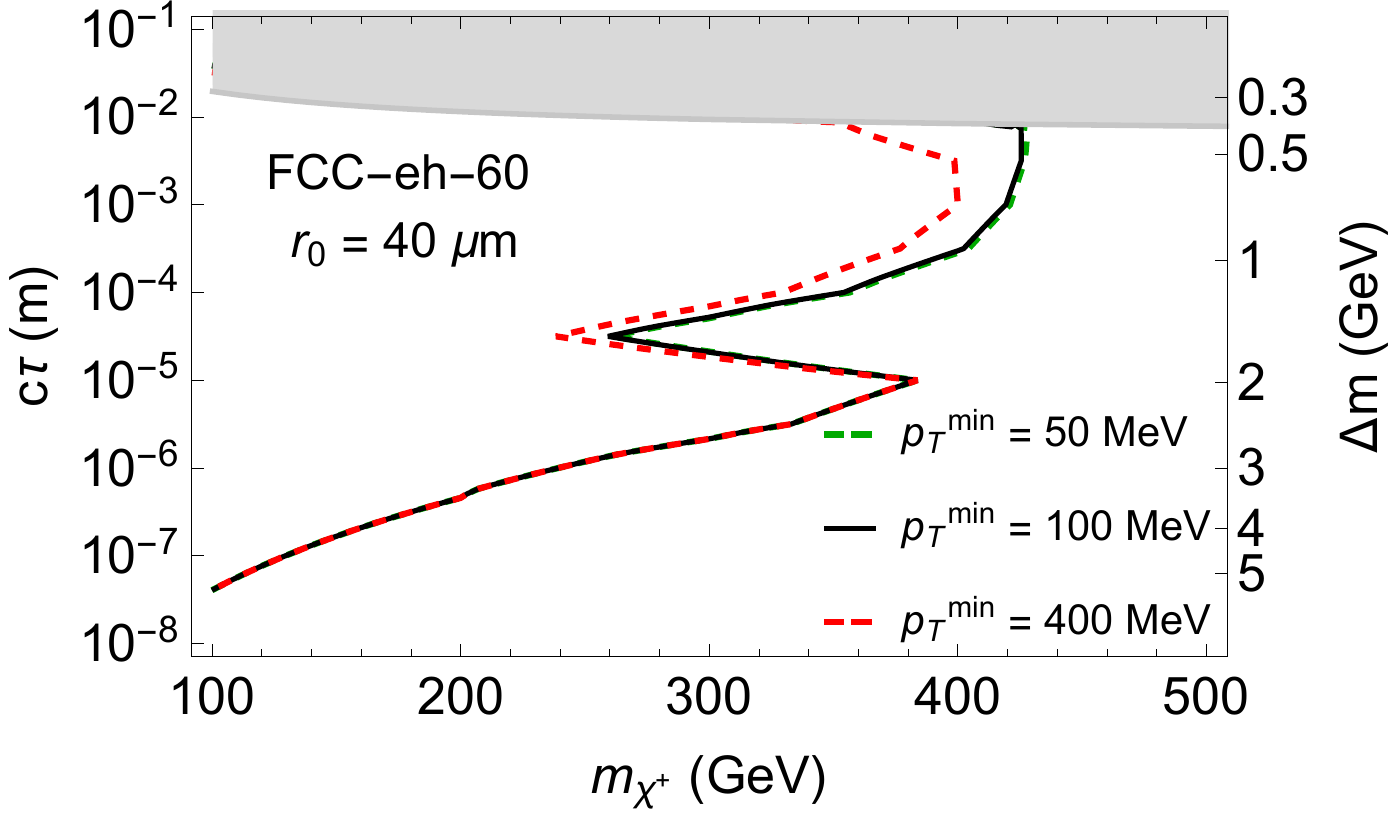}
\\
\includegraphics[width=8cm]{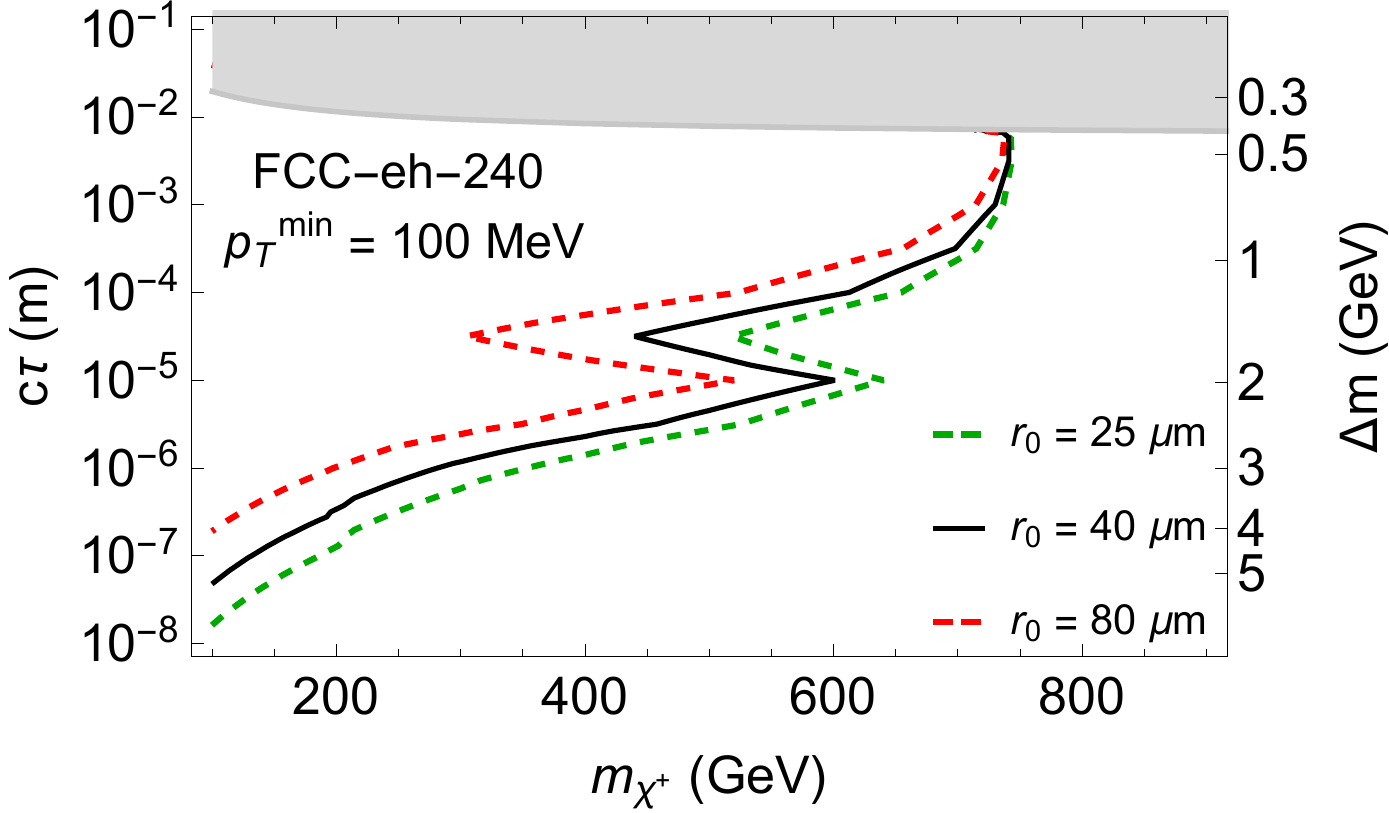}
&
\includegraphics[width=8cm]{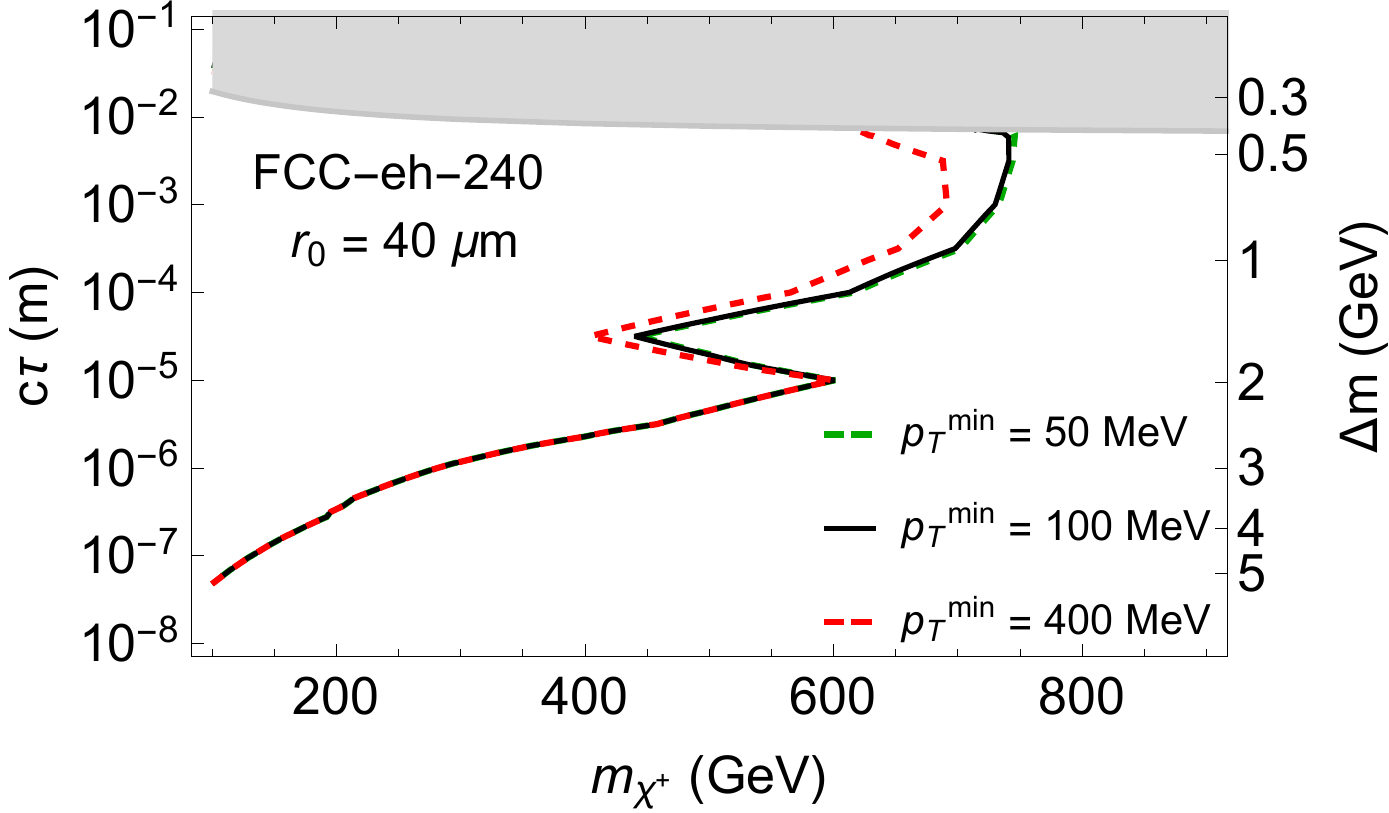}
\end{tabular}
\end{center}
\caption{Reach dependence on $r_0$ and $p_T^\mathrm{min}$. All plots assume $1\iab$ of data, $\mu > 0$, and the most optimistic estimate for event yield given hadronization and displaced jet reconstruction uncertainties.}
\label{f.varypTminr0}
\end{figure*}

\subsubsection*{FCC-eh}

We repeat the above analysis for the FCC-eh scenarios. We assume the same detector dimensions, triggers, and thresholds. The kinematic rejection of $\tau$ backgrounds improves, with rejections in the range of $10^{-4} - 10^{-3}$ ($10^{-5} - 10^{-4}$) for one (two) $\tau$ events, more than offsetting the modest growth in $\tau$-cross section, which is 2.1 (0.8) pb at the FCC-eh with a 60 GeV electron beam, and 4.4 (1.1) pb with a 240 GeV electron beam. 

Figs.~\ref{f.higgsinoFCCeh60} and \ref{f.higgsinoFCCeh240} show the number of observed events with at least 1 or 2 LLPs at the FCC-eh (60) and FCC-eh (240). 
We recall that we here consider benchmark luminosities of 1 and 10 $\iab$.
For the latter, we show contours of 300 and 30 events instead of 100 and 10 to estimate sensitivity. This roughly accounts for the $\sqrt{10}$ larger number of signal events required to stand out against the same background cross section with a factor of 10 higher luminosity. However, we also show contours for 10 events, in the event that background rejection is very good and sensitivity scales more linearly with luminosity. We emphasize that the FCC-eh (240) with $10\iab$ of luminosity may be able to probe the 1.1 TeV thermal Higgsino DM relic at lifetimes much shorter than FCC-hh disappearing track searches. 
Furthermore, this reach is theoretically very robust since LLP tagging efficiency at $\mathcal{O}(\mathrm{mm})$ lifetime is excellent at $e^- p$ colliders.

We note that an $\mathcal{O}(1)$ pile-up may become  relevant at higher beam energies and luminosities.
A detailed discussion is beyond our scope, but we expect that single displaced charged particles should be kinematically clearly distinguishable from a second high-energy primary vertex. Furthermore, given the sizable longitudinal extent of the interaction region, sensitivity at short lifetimes would not be affected by requiring the impact parameter or DV distance from the PV to be much less than the beam spot length. This would further reject pile-up vertices, which are more evenly distributed along the beam axis. 
While a more thorough investigation is certainly required, we expect our results to be fairly robust against these modest levels of pile-up, especially for the search requiring 2 observed LLPs.

\subsubsection*{Impact of track resolution and energy thresholds}

It is important to determine to what extent the specifications of the detector, like energy thresholds and tracking resolution, affect BSM reach. In \fref{varypTminr0} we show how reach is modified if we deviate from our benchmark assumptions of $p_T^\mathrm{min} = 100 \mev$ as the minimum threshold for single track reconstruction and $r_0^\mathrm{min} = 40 \mu\mathrm{m}$ as the minimum spatial separation for LLP tagging. 

Our results are fairly robust with respect to variation in these two thresholds. Changing the tracking resolution ($r_0^\mathrm{min}$) unsurprisingly has noticeable effect on reach at the lowest lifetimes, but does not affect mass reach at the larger lifetimes. Conversely, the $p_T^\mathrm{min}$ threshold has no effect on reach at short lifetimes (where mass splitting is larger, leading the single charged particles to always pass the threshold). At large lifetimes the benchmark threshold of 100 MeV is very close to optimal, with improvements for 50 MeV being very minimal. On the other hand, assuming a much worse threshold of 400 MeV would modestly affect mass reach, which would make it even harder to reach the $m_\chi = 1.1 \tev$ goal corresponding to thermal Higgsino dark matter. This provides significant motivation to aim for single track reconstruction thresholds at the $\sim 100 \mev$ level when finalizing detector design.

\subsubsection*{Discussion and comparison}

Our projected LHeC sensitivity for Higgsinos is competitive in mass reach to the monojet projections for the HL-LHC, being sensitive to masses around 200 GeV for the longest theoretically motivated lifetimes. The LHeC search has the crucial advantage of actually observing the charged Higgsino parent of the invisible final state. Proposed disappearing track searches at the HL-LHC may probe higher masses for the longest lifetimes, but lose sensitivity at shorter lifetimes. 
By comparison, the LHeC search is sensitive to lifetimes as short as microseconds. 
It is important to note that the mass reach of $e^- p$ colliders is much more robust than the disappearing track projections, since the former are not exponentially sensitive to uncertainties in the Higgsino velocity distribution. 
While similar lifetime sensitivities may be possible at lepton colliders, only the highest energy proposals would have comparable center-of-mass energy. 

The direct collider sensitivities are complementary to the sensitivity of dark matter direct detection experiments, which cover larger mass splittings (shorter lifetimes), and indirect detection constraints. However, these bounds are model-dependent and rely on cosmological assumptions. 
In the event of a positive dark matter signal, $e^- p$ colliders would play a crucial role in determining the nature of the dark matter candidate.

The mass reach of the FCC-eh is obviously much greater than for the LHeC. Reaching the thermal Higgsino DM mass of $\sim 1.1 \tev$ is challenging and would require a high luminosity high energy FCC-eh scenario as shown in \fref{higgsinoFCCeh240} (left). However, in all cases the sensitivity to short decay lengths, possibly much less than a single micron, far exceeds what the FCC-hh can accomplish with disappearing track searches, making the FCC-eh coverage crucial in probing the full range of possible Higgsino scenarios.

\section{LLP Production in Exotic Higgs Decays}
\label{s.higgs}

The Higgsino analysis of the previous section demonstrates that $e^- p$ colliders have unique capabilities to detect LLPs which decay due to almost-degenerate masses into extremely soft SM final states with very short lifetimes. However, the excellent tracking resolution, clean environment and longitudinal boost of the collision center-of-mass frame also has significant advantages for detecting LLPs with somewhat higher energy final states.

Exotic Higgs decays are strongly motivated on general theoretical grounds, see e.g. ref.~\cite{Curtin:2013fra}:  the small SM Higgs width allows even small BSM couplings to lead to sizable exotic Higgs branching fractions, and the low dimensionality of the gauge- and Lorentz-singlet $|H|^2$ portal operator allows it to couple to any BSM sector via a low-dimensional term in the Lagrangian, making sizable couplings generic.

We consider exotic Higgs decays into a pair of BSM LLPs $X$. The exotic branching fraction $\mathrm{Br}(h \to X X)$  and the LLP lifetime $c \tau$ are both essentially free parameters. We focus on LLP masses of order 10 GeV to demonstrate that $e^- p$ colliders also offer crucial advantages to LLPs without soft decay products. 
This simplified model represents many highly motivated theoretical scenarios, including Neutral Naturalness~\cite{Craig:2015pha} and general Hidden Valleys~\cite{Strassler:2006im,Strassler:2006ri,Strassler:2006qa,Han:2007ae,Strassler:2008bv,Strassler:2008fv}, where the LLPs are hadrons of the hidden sector produced via the Higgs portal.

\subsubsection*{Analysis strategy}

We assume $X$ decays to at least two charged particles with energies above $p_T$ detection threshold to uniquely identify a DV for the LLP decay.  The analysis proceeds along very similar lines as the Higgsino case: VBF Higgs production at $e^- p$ colliders, see \fref{feynman} (right), is simulated to lowest order in MadGraph, with cross sections 0.1, 0.34, 1.05 pb at the LHeC, FCC-eh (60) and FCC-eh (240) respectively. The search strategy is also the same, shown in \fref{displaced}, but now we are dealing exclusively with displaced vertices (C), which we assume are detected with an efficiency of 100\% as long as the final states hit the tracker and the LLP decays at a distance $r_{min}$ away from the primary vertex, which is again identified by the associated jet which passed the trigger.

Our search requires, in addition to the triggering jet, a single tagged DV consistent with the decay of an LLP of mass a few GeV or above.\footnote{The sensitivity at longer lifetimes $\gtrsim 10 \mathrm{cm}$ depends more on the detector geometry, as well as the final state, so for simplicity we simply assumed that DVs within 1 meter of the PV can be reconstructed.} Requiring an invariant mass of the DV above about 10 GeV and making use of the known Higgs mass (for both DVs decaying in the detector)
 efficiently rejects background events from $\tau$ leptons or other sources.
 We note that this search can be generalized for other resonant LLP production processes.

\begin{figure}
\begin{center}
\includegraphics[width=0.5\textwidth]{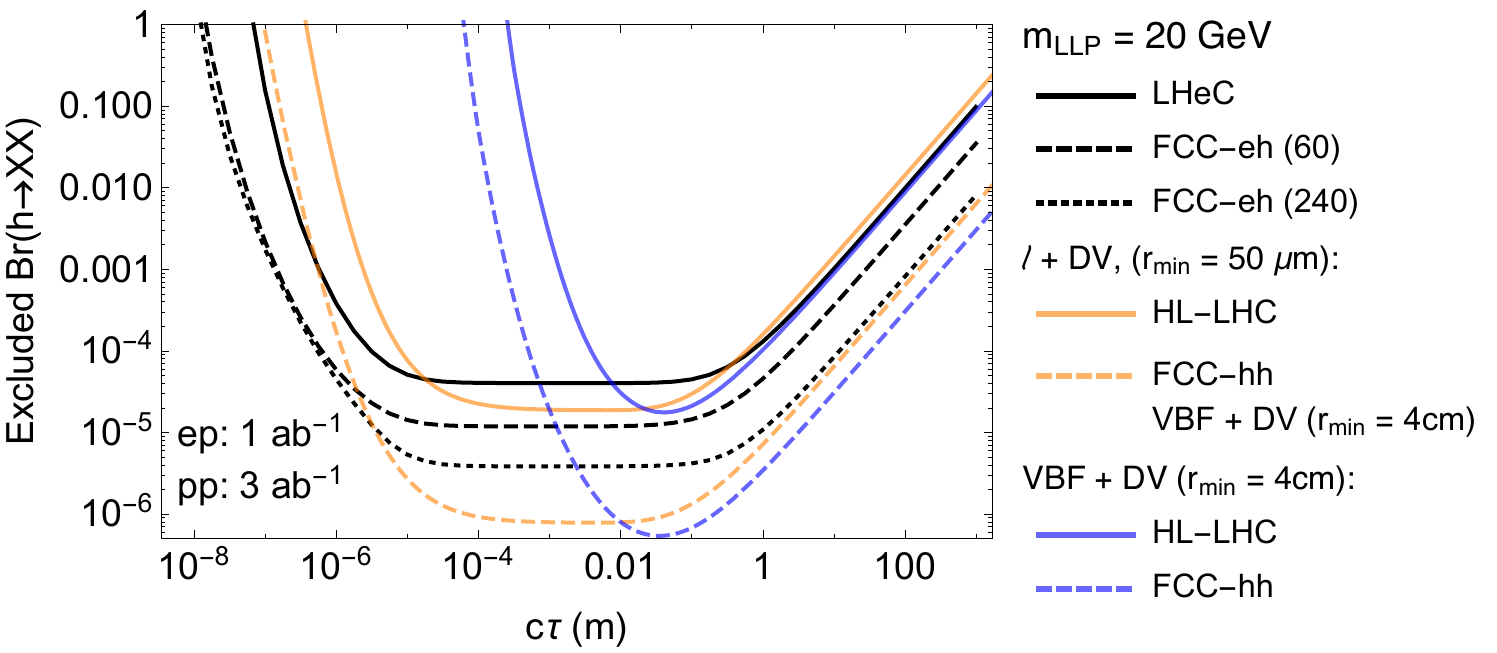}
\end{center}
\caption{
Projected exclusion limits on exotic Higgs decay branching fraction to LLPs $X$ as a function of lifetime $c \tau$ for the LHeC, FCC-eh (60) and FCC-eh (240) with $1\iab$ of data. The excluded branching ratio scales linearly with luminosity under the assumption of no background.
The LLP mass in the plot is $20 \gev$, but for different masses the curves shift in $c \tau$ roughly by a factor of $m_\mathrm{LLP}/(20 \gev)$. For comparison, we show a somewhat realistic estimate for the sensitivity of $pp$ colliders with $3\iab$ and without background (blue), as well as a very optimistic estimate which assumes extremely short-lived LLP reconstruction (orange), from \cite{Curtin:2015fna}. 
}
\label{f.higgslimits}
\end{figure}

\subsubsection*{Results and discussion}
We show the resulting sensitivity in \fref{higgslimits}, with the exclusion sensitivity of 4 expected events.
From the figure we see that $e^-p$ colliders can probe LLP production in exotic Higgs decays with decay lengths below a micron, due to the lifetime-enhancing longitudinal boost and excellent tracking in a clean environment. 

For comparison, we show estimates of the HL-LHC and FCC-hh sensitivity to LLPs produced in exotic Higgs decays~\cite{Curtin:2015fna}. A somewhat realistic estimate assumes triggering on Higgs production from VBF\footnote{This reach estimate would be very similar if the search triggered on leptons from associated production instead of VBF.} and requiring a single DV displaced more than 3cm from the beamline is enough to eliminate backgrounds (blue curves). A much more optimistic estimate (orange curves) assumes a search triggering on a single high-$p_T$ lepton from associated Higgs Boson production and requiring a single DV with displacement as low as $50 \mu m$ can be performed with no backgrounds. It is still unclear whether this optimistic search can be realized at $pp$ colliders.

The sensitivity achievable at the LHeC (FCC-eh) reaches much shorter lifetimes than either projection for the HL-LHC (FCC-hh), especially for the more conservative $pp$ projections. 
This is especially significant since the optimistic search of \cite{Curtin:2015fna} was required to cover well-motivated parts of Neutral Naturalness parameter space where the hidden hadrons are very short-lived. 
Furthermore, the estimated sensitivity of $e^- p$ colliders at short lifetimes is more robust than that of $pp$ colliders, where those searches have to contend with much higher levels of background and pile-up.

\section{Conclusion}
\label{s.conclusion}

Electron-proton colliders are more commonly associated with DIS studies of the proton than with BSM searches. However, their high center-of-mass energy compared to lepton colliders but clean environment compared to hadron colliders lets them play a unique role in probing a variety of important BSM signals. 

Diverse BSM states can be produced in VBF processes, which also ensures triggering and identification of the primary vertex. Any BSM state which looks like hadronic background in the high-energy, high-rate environment of hadron colliders can likely be much better identified and studied in $e^- p$ collisions. A prime example of such BSM scenarios are LLPs which decay with short lifetime $(\lesssim$ mm) and/or a small mass splitting $(\lesssim \gev)$ which can arise from compressed spectra. To demonstrate this, we studied searches for pure Higgsinos and exotic Higgs decays to LLPs. In both cases, proposed $e^- p$ colliders probe new and important regions of parameter space inaccessible to other experiments. 
Our most optimistic FCC-eh scenarios could produce and reconstruct the 1.1 TeV  thermal Higgsino dark matter relic. 
It is also important to point out that in both BSM scenarios, the $e^- p$ collider reach is more robust than the $pp$ projections.

We used LHeC and FCC-eh proposals as our benchmarks, but took some liberties in exploring higher luminosities and higher energies to show what kind of physics reach may be possible. 
In that light, our results can serve to guide the detailed design of such a future machine, whether it is built as an add-on to the CERN LHC, CERN FCC-hh, or at the SppC. 
Similarly, we found that the reconstruction of soft LLP final states with high tracking resolution ($\lesssim 10 \mu m$), single track reconstruction thresholds of  $\sim 100 \mev$ and very low pile-up are necessary conditions for this unique BSM sensitivity, and should be a high priority in the design.

We demonstrated that $e^- p$ colliders have unique sensitivity to BSM signals, in particular LLPs with soft final states or very short lifetimes. 
Further study is needed to identify other BSM scenarios to which these machines could be uniquely sensitive, but our results suggest that difficult final states may be a particularly fruitful avenue of exploration. 
There may be other diverse classes of signals that can be effectively probed. 
This adds significant motivation for the construction of future $e^- p$ colliders. 
Together with the invaluable proton PDF data, as well as precision measurements of EW parameters, top quark couplings and Higgs couplings, 
our results make clear that adding a DIS program to a $pp$ collider is necessary to fully exploit its discovery potential for new physics.

\acknowledgments
We thank Raman Sundrum for helpful conversations. 
We especially thank Max Klein, Uta Klein, Peter Kostka, Monica d'Onofrio, Georges Azuelos, Sho Iwamoto for useful discussions about $e^-p$ signal simulation, backgrounds, and detector capabilities.
D.C., and K.D. are supported by National Science Foundation grant No. PHY-1620074 and the Maryland Center for Fundamental Physics. 
O.F.\ acknowledges support from the ``Fund for promoting young academic talent'' from the University of Basel under the internal reference number DPA2354 and has received funding from the European Unions Horizon 2020 research and innovation program under the Marie Sklodowska-Curie grant agreement No 674896 (Elusives).

\bibliography{ep_collider_LLP}

\end{document}